\documentclass[pdflatex,sn-nature]{sn-jnl}% Style for submissions to Nature Portfolio journals
\usepackage{graphicx}%
\usepackage{multirow}%
\usepackage{amsmath,amssymb,amsfonts}%
\usepackage{amsthm}%
\usepackage{mathrsfs}%
\usepackage[title]{appendix}%
\usepackage{xcolor}%
\usepackage{textcomp}%
\usepackage{manyfoot}%
\usepackage{booktabs}%
\usepackage{algorithm}%
\usepackage{algorithmicx}%
\usepackage{algpseudocode}%
\usepackage{listings}%
\usepackage{dsfont}
\usepackage{anyfontsize}
\usepackage{subcaption}
\usepackage{makecell}

\theoremstyle{thmstyleone}%
\theoremstyle{thmstyletwo}%

\theoremstyle{thmstylethree}%

\begin{document}

\title[Article Title]{Multimodal Deep Learning for Uncertainty-Aware Radiation Pneumonitis Risk Prediction}

\author[1]{\fnm{Jin} \sur{Yang}}

\author[1]{\fnm{Tian} \sur{Liu}}

\author[1]{\fnm{Jing} \sur{Wang}} 

\author[2]{\fnm{Sooyun Caroline} \sur{Tavolacci}}

\author[1]{\fnm{Benjamin} \sur{Tran}}

\author[3]{\fnm{Rajwanth} \sur{Veluswamy}}

\author[1]{\fnm{Kenneth E.} \sur{Rosenzweig}}

\author[1]{\fnm{Julie} \sur{Bloom}}

\author[1]{\fnm{Robert M.} \sur{Samstein}}

\author*[1]{\fnm{Ming} \sur{Chao}}\email{ming.chao@mountsinai.org}

\affil[1]{\orgdiv{Department of Radiation Oncology}, \orgname{Icahn School of Medicine at Mount Sinai}, \orgaddress{\city{New York}, \state{NY} \postcode{10029}, \country{USA}}}

\affil[2]{\orgdiv{Center for Thoracic Oncology}, \orgname{Icahn School of Medicine at Mount Sinai}, \orgaddress{\city{New York}, \state{NY} \postcode{10029}, \country{USA}}}

\affil[3]{\orgdiv{Department of Hematology and Oncology}, \orgname{NYU Langone Health Perlmutter Cancer Center, NYU Grossman Shool of Medicine}, \orgaddress{\city{New York}, \state{NY} \postcode{10016}, \country{USA}}}

\abstract{Radiation pneumonitis (RP) is a potentially serious complication of thoracic radiotherapy, making reliable risk prediction important for identifying high-risk patients and guiding effective toxicity management. Although deep learning models can predict RP risk, uncertainty quantification remains underdeveloped in radiation toxicity assessment. We developed a Multimodal Bayesian Diffusion Transformer (MM-DiT) that integrates pretreatment computed tomography (CT) and three-dimensional planning dose distributions for RP risk prediction and uncertainty quantification. MM-DiT combines self-supervised multimodal representation learning with Bayesian classification to estimate individual risk and characterize aleatoric, epistemic, and label uncertainty. In independent Mount Sinai Health System (MSHS) and RTOG cohorts, MM-DiT achieved areas under the receiver operating characteristic curve (AUCs) of $0.89$ and $0.86$, respectively, with expected calibration errors of $0.02$ and $0.03$. Incorporating post-treatment CT increased AUC to $0.95$ in MSHS and reduced aleatoric and epistemic uncertainty by approximately $40-45\%$ and $40\%$, respectively. These findings establish a framework for uncertainty-aware radiation toxicity prediction that complements risk estimates with information about predictive reliability. By integrating multimodal learning and uncertainty quantification, MM-DiT may advance more interpretable and trustworthy AI-assisted toxicity assessment and support individualized radiotherapy planning.}


\maketitle

\section{Introduction}
\label{sec1}
Radiation pneumonitis (RP) is a potentially serious complication of thoracic radiation therapy (RT) that can limit radiation dose escalation needed for effective tumor control \cite{kaesmann2020radiation,maddali2024optimal,de2025diagnosis}. Developing within weeks to months after treatment, RP ranges from mild respiratory symptoms to severe pulmonary inflammation and, rarely, respiratory failure \cite{kraus2024dosiomics,weisman2025evaluation,xiao2025early}. Clinically significant RP can lead to hospitalization, impair quality of life, progress to pulmonary fibrosis, and disrupt subsequent systemic therapy.

Current clinical strategies to mitigate RP rely primarily on pretreatment lung dose constraints, including recommendations from the Quantitative Analyses of Normal Tissue Effects in the Clinic (QUANTEC) for mean lung dose and the percentage of lung receiving specified doses \cite{marks2010radiation}. Risk is also commonly estimated using dose-volume-histogram (DVH)-based normal tissue complication probability models such as the Lyman-Kutcher-Burman (LKB) model \cite{lyman1985complication,kutcher1989calculation,kutcher1991histogram}. However, these approaches suffer from critical structural limitations. DVH metrics summarize three-dimensional (3D) dose distributions into 1D summary statistics, discarding crucial spatial context and regional tissue heterogeneity. Furthermore, conventional DVH-based models fail to integrate patient-specific anatomical variations captured on planning computed tomography (CT) scans, nor do they account for non-dosimetric drivers such as baseline pulmonary function, tumor geometry, and individual radiosensitivity \cite{van2020key}. Consequently, patients with similar DVH metrics may experience markedly different clinical outcomes \cite{he2024quantifying,liu2025relative}.

To address these limitations, machine learning methods, including multimodal approaches that integrate imaging, dosimetric, and clinical information, have emerged as promising tools for individualized RP prediction \cite{valdes2016using,krafft2018utility,yu2019machine,luna2019predicting,hirose2020radiomic,feng2024improvement,gadsby2025impact}. By learning nonlinear patterns and interactions across heterogeneous data sources, these approaches may better capture the complex relationships among pulmonary anatomy, radiation exposure, and patient-specific susceptibility \cite{cui2021integrating,jiang2021dosimetric,jiao2025integrating,reuter2026prediction}. Nevertheless, almost all existing RP models output a single point-estimate risk score without quantifying the confidence surrounding that prediction. This creates a significant barrier to clinical adoption: strong population-level performance does not guarantee individual reliability, particularly when a patient presents with atypical anatomy or clinical characteristics underrepresented in the training data.

Although uncertainty quantification has increasingly been recognized as essential for trustworthy artificial intelligence (AI) approaches for medical prediction \cite{kurz2022uncertainty,loftus2022uncertainty,huang2024review,stutz2025evaluating,guennoun2026segmenting,wang2025multimodal}, its application to RT-induced toxicity assessment remains largely unexplored \cite{wahid2024artificial,dunger2025uncertainties}. This gap is particularly important for RP prediction, where uncertainty arises from distinct sources: aleatoric uncertainty, reflecting inherent variability in patient-specific responses to radiation \cite{nair2020exploring,herzog2020integrating}; epistemic uncertainty, reflecting limitations in model knowledge due to insufficient or underrepresented training data \cite{kendall2017uncertainties,herzog2020integrating}; and label uncertainty, arising from ambiguity or inconsistency in toxicity assessment and annotations \cite{yang2023parametrical,wei2024deep}. 
Distinguishing these sources may help clarify whether uncertainty reflects biological variability, limited model knowledge, or ambiguity in the observed outcome. A clinically informative framework should therefore jointly estimate individual RP risk and characterize complementary sources of predictive uncertainty to better support patient-specific risk stratification and toxicity management.

\begin{figure*}[!ht]
	\centering
	\includegraphics[width=0.75\linewidth]{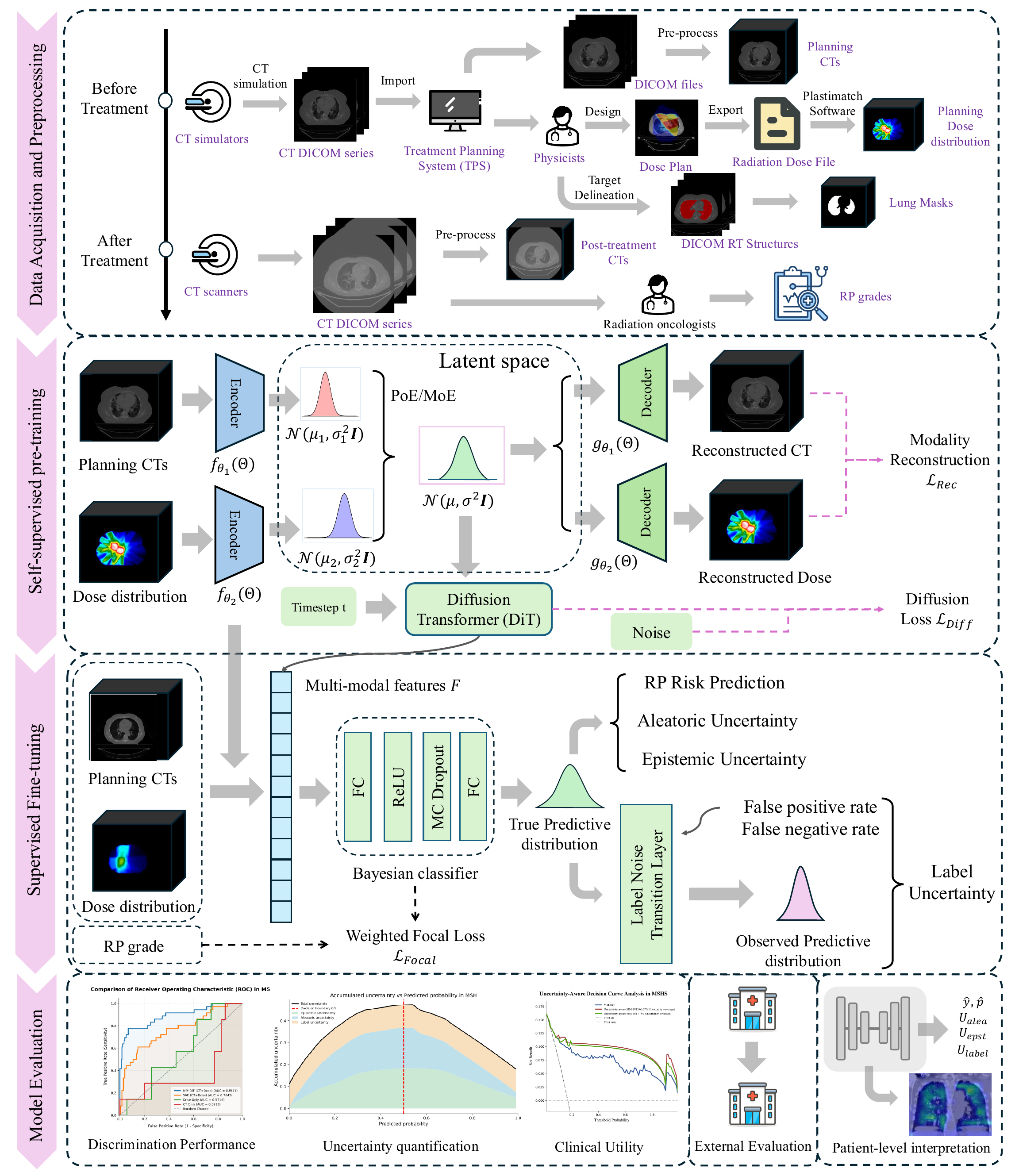} \\
	\caption{\textbf{Multimodal Bayesian Diffusion Transformer framework for radiation pneumonitis risk prediction and uncertainty quantification}. Pretreatment CT images were collected and imported into the Treatment Planning System (TPS) as DICOM series, while treatment plans and corresponding dose files were generated by clinical physicists on the CT images. Dose files were then exported and processed using Plastimatch to generate 3D planning dose distributions. Lung masks were selected from radiation therapy structures and used to facilitate preprocessing and anatomical region extraction. Pretreatment CT and 3D planning dose distributions were independently encoded using two shared-architecture encoders, followed by multimodal fusion and diffusion transformer-based representation learning. The encoders and diffusion transformer were first optimized through self-supervised pre-training and subsequently fine-tuned with a Bayesian classifier for RP prediction. During inference, the Bayesian classifier estimates RP risk and predictive uncertainty, including aleatoric and epistemic uncertainty, while a learnable label-noise transition layer models uncertainty arising from imperfect toxicity labels. MM-DiT was evaluated for discrimination, uncertainty quantification, and clinical utility using internal evaluation, and for generalizability using external evaluation.}
	\label{fig:model}
\end{figure*}

To address these needs, we developed a Multimodal Bayesian Diffusion Transformer (MM-DiT) for uncertainty-aware RP risk prediction (Figure \ref{fig:model}). MM-DiT addresses two complementary challenges: learning spatial relationships between pulmonary anatomy and radiation dose, and quantifying uncertainty in the resulting risk estimate. Imaging–dosimetric modeling provides a complementary approach to characterizing the spatial relationships between pulmonary anatomy and radiation exposure, although integrating these data with clinical factors may be necessary to capture the full spectrum of patient-specific RP susceptibility. Two encoders with a shared architecture independently extract latent representations from pre-treatment CT images and 3D planning dose distributions, preserving complementary anatomical and dosimetric information that conventional DVH metrics cannot fully represent. These representations are fused and further integrated by a diffusion transformer to learn a joint features capturing interactions between patient anatomy and radiation dose. To reduce reliance on limited and potentially noisy toxicity labels, the encoders and diffusion transformer are first optimized through self-supervised pre-training to learn robust and generalizable multimodal representations without RP outcome labels. The pre-trained components are then integrated with a Bayesian classifier and jointly optimized through supervised fine-tuning, enabling the learned representations to be adapted specifically for RP prediction while retaining the benefits of multimodal pre-training. During inference, Bayesian modeling produces a predictive distribution from which individual RP risk and aleatoric and epistemic uncertainty are derived. A learnable label-noise transition layer additionally models label uncertainty associated with imperfect toxicity annotations. 

We evaluated MM-DiT in two independent cohorts, RTOG 0617 (a multi-institutional trial dataset) and the Mount Sinai Health System (MSHS) clinical cohort, assessing discrimination, calibration, and predictive uncertainty. Decision curve analysis (DCA) was used to examine potential clinical utility across different risk thresholds. To investigate how additional patient-specific information influences prediction, we further incorporated post-treatment CT in the MSHS cohort and assessed changes in RP risk estimates and their associated uncertainty. Together, these analyses evaluate not only predictive performance but also whether uncertainty estimates provide complementary information about individual prediction reliability and how that information changes when additional treatment-response data become available.

\section{Results}
We first evaluated MM-DiT for pretreatment RP risk prediction using pretreatment CT images and 3D planning radiation dose distributions. Section \ref{sec21} presents the predictive performance and calibration of MM-DiT, including comparisons with existing methods and unimodal models to demonstrate its improved discriminative performance and the value of integrating multimodal information for RP risk prediction. Section \ref{sec22} evaluates the predictive uncertainty quantified by MM-DiT and examines its potential clinical utility. Section \ref{sec23} investigates the impact of incorporating post-treatment CT images alongside pretreatment CT and planning dose distributions, demonstrating improvements in both RP risk prediction and uncertainty quantification. Section \ref{sec24} presents an external evaluation of MM-DiT across independent cohorts to assess its generalizability. Finally, Section \ref{sec25} presents three representative cases illustrating how MM-DiT can support individualized RP risk assessment and uncertainty-aware clinical decision-making.

\subsection{Performance and Calibration of RP Risk Prediction}
\label{sec21}
\begin{figure*}[!t]
    \centering
    \begin{subfigure}[t]{0.49\linewidth}
        \centering
        \includegraphics[width=\linewidth]{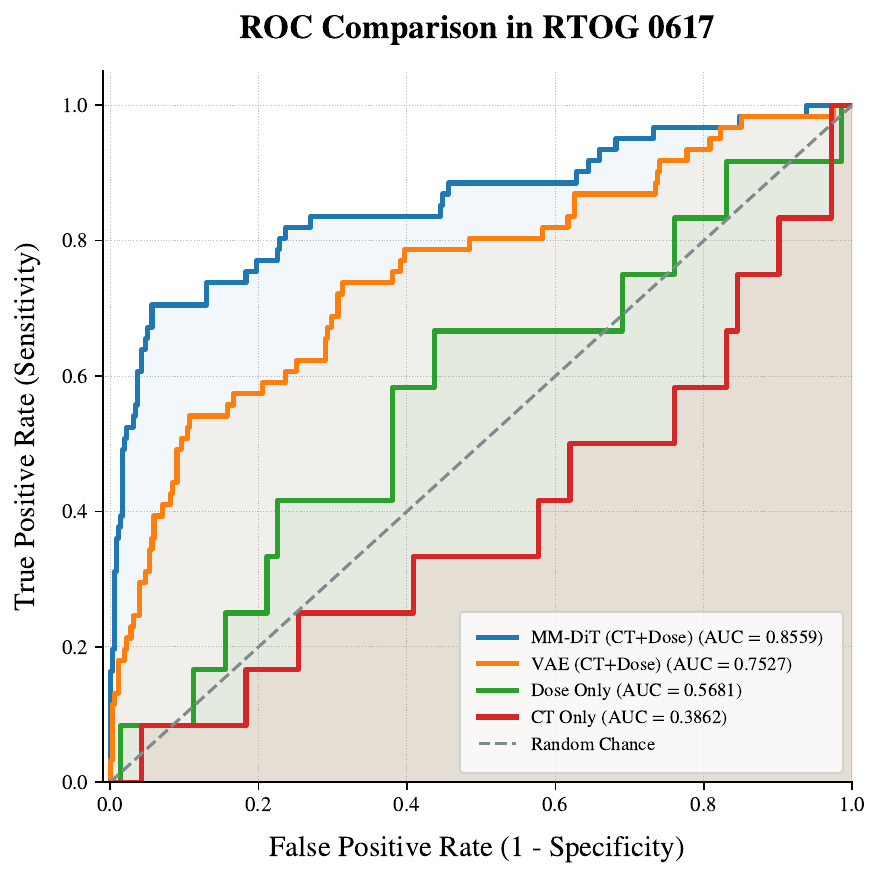}
        \caption{ROC curves comparison on the RTOG 0617 cohort.}
        \label{fig:roc_rtog2}
    \end{subfigure}
    \hfill
    \begin{subfigure}[t]{0.49\linewidth}
        \centering
        \includegraphics[width=\linewidth]{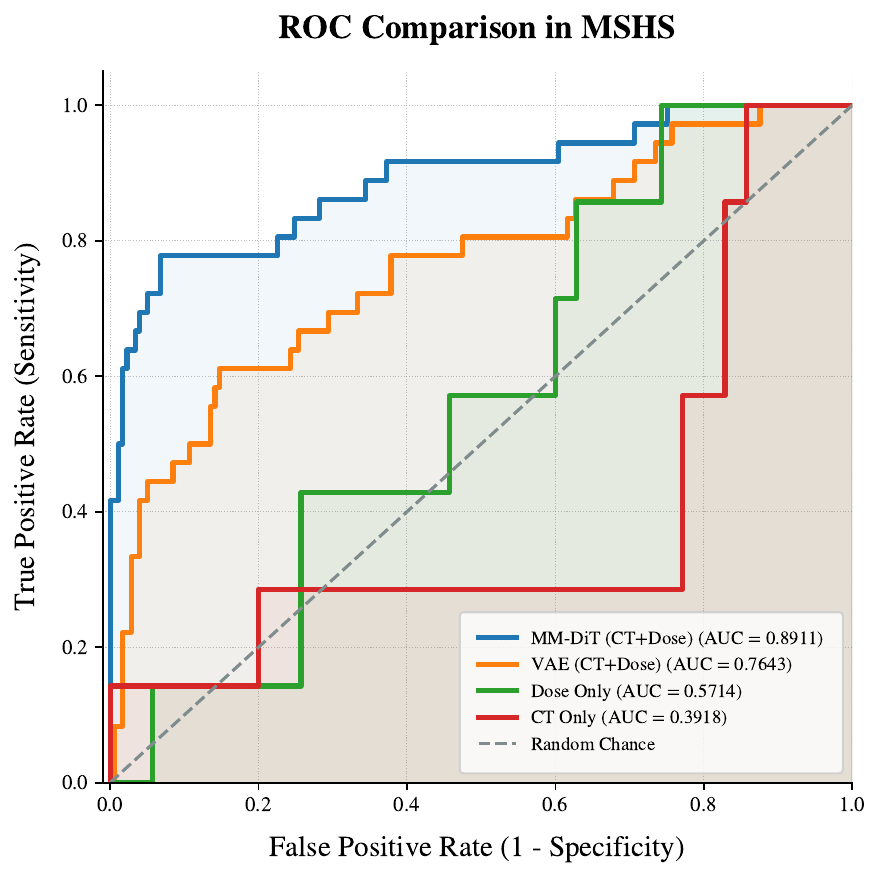}
        \caption{ROC curves comparison on the MSHS cohort.}
        \label{fig:roc_msh2}
    \end{subfigure}
    \caption{Comparison of receiver operating characteristic (ROC) curves for the evaluated radiation pneumonitis prediction models in the (a) RTOG 0617 and (b) MSHS cohorts.}
    \label{fig:roc}
\end{figure*}

Table~\ref{tab1} summarizes the RP prediction performance of the evaluated models in the RTOG 0617 and MSHS cohorts. First, we evaluated the contribution of multimodal information by comparing models using pretreatment planning CT or 3D radiation dose distributions alone with models jointly incorporating both modalities. The MM-DiT configurations achieved AUCs of $0.824$ - $0.856$ in RTOG 0617 and $0.854$ - $0.891$ in MSHS, demonstrating the complementary value of anatomical and dosimetric information for RP risk prediction (Table~\ref{tab1} and Figure \ref{fig:roc}). In contrast, the unimodal models exhibited limited discriminative performance in both cohorts, with AUCs of $0.386$ and $0.568$ for CT-only and dose-only models, respectively, in RTOG 0617, and $0.392$ and $0.571$ in MSHS. The improvement was also reflected in sensitivity ($0.459$ - $0.639$ vs. $0.082$ of CT- or dose-only models), with multimodal models generally identifying a substantially greater proportion of RP cases than the corresponding unimodal models. These results support the importance of jointly modeling anatomical and dosimetric information rather than relying on either modality alone.

Second, we compared MM-DiT with the multimodal variational autoencoder (MM-VAE) models, demonstrating its superior performance. MM-DiT achieved the highest AUC of $0.856$ in RTOG 0617 and $0.891$ in MSHS (Table~\ref{tab1}). Additionally, the MM-DiT model with product-of-experts (PoE) fusion achieved the highest AUC and accuracy in both cohorts, outperforming configurations using ViT-based encoders, mixture-of-experts (MoE) fusion. Additionally, the MM-DiT with different configurations outperformed MM-VAE across both cohorts. These results demonstrate the advantage of diffusion-transformer-based multimodal representation learning over the VAE-based multimodal approach for RP risk prediction.

Finally, we compared MM-DiT with its deterministic variants to assess the impact of Bayesian inference on predictive performance. The Bayesian MM-DiT with PoE achieved AUCs of $0.856$ and $0.891$ in RTOG 0617 and MSHS, respectively, comparable to $0.851$ and $0.885$ for the corresponding deterministic models. These results indicate that Bayesian uncertainty quantification can be incorporated without compromising the predictive performance of MM-DiT. Notably, the Bayesian MM-DiT with PoE achieved the highest AUC and accuracy among all evaluated models.

\begin{table*}[!t]
\centering
\caption{Performance comparison of unimodal and multimodal models for radiation pneumonitis prediction using pre-treatment planning CT and 3D radiation dose distributions in the RTOG 0617 and MSHS cohorts. Models were evaluated using the area under the receiver operating characteristic curve (AUC), accuracy, specificity, and sensitivity. \textbf{Bold} indicates the best performance for each metric across the evaluated models. $p_{\mathrm{adj}}$ denotes the false discovery rate-adjusted $p$-value.}
\label{tab1}
\resizebox{\textwidth}{!}{
\begin{tabular}{c|c|c|ccccc|ccccc}
\toprule
\multirow{2}{*}{Modalities} & \multirow{2}{*}{Models} & \multirow{2}{*}{Fusion} &  \multicolumn{5}{c|}{RTOG 0617} &  \multicolumn{5}{c}{MSHS} \\
\cmidrule(lr){4-13}
& & & AUC ($95\%$CI) & Accuracy ($95\%$CI) & Specificity ($95\%$CI) & Sensitivity ($95\%$CI) & $p_{\tiny adj}$ & AUC ($95\%$CI) & Accuracy ($95\%$CI) & Specificity ($95\%$CI) & Sensitivity ($95\%$CI) & $p_{\tiny adj}$ \\
\midrule
Planning CT & CNN & - & 38.62 (31.44, 45.80) & 79.33 (75.18, 82.94) & 91.55 (88.19, 94.02) & 8.20 (3.55, 17.79)  & $<0.001$ & 39.18 (29.61, 48.75) & 72.77 (66.42, 78.30) & 84.75 (78.71, 89.30) & 13.89 (6.08, 28.66) & $<0.001$ \\
3D Dose     & CNN & - & 56.81 (48.80, 64.82) & 85.34 (81.61, 88.41) & 98.59 (96.75, 99.40) & 8.20 (3.55, 17.79)  & $<0.001$ & 57.14 (46.58, 67.70) & 68.08 (61.54, 73.97) & 73.44 (66.49, 79.41) & 41.67 (27.14, 57.80) & $<0.001$ \\
Planning CT+3D Dose & MM-VAE & Concatenation & 75.27 (67.86, 81.85) & 86.54 (83.41, 89.66) & 98.87 (97.53, 99.72) & 14.75 (6.67, 23.81) & $<0.001$ & 76.43 (66.74, 85.19) & 84.51 (79.81, 89.20) & 92.66 (88.71, 96.11) & 44.44 (28.20, 60.00) & $<0.001$ \\
Planning CT+3D Dose & MM-VAE & MoE           & 80.36 (73.42, 86.66) & 87.26 (83.89, 90.38) & \textbf{99.72} (99.14, 100.00) & 14.75 (6.25, 24.24) & $<0.001$ & 83.90 (75.60, 91.51) & 87.32 (83.10, 91.55) & 93.22 (89.33, 96.61) & 58.33 (41.67, 75.00) & $<0.001$ \\
Planning CT+3D Dose & MM-VAE & PoE           & 81.18 (74.87, 87.05) & 88.22 (85.10, 91.35) & 97.75 (96.06, 99.16) & 32.79 (21.05, 44.83) & $<0.001$ & 84.95 (75.12, 91.44) & 87.79 (83.57, 92.02) & 97.18 (94.50, 99.43) & 41.67 (25.71, 58.33) & $<0.001$ \\
Planning CT+3D Dose & ViT MM-DiT & MoE & 82.37 (75.50, 89.14) & 88.22 (84.86, 91.35) & 95.49 (93.22, 97.47) & 45.90 (33.85, 59.38) & $<0.01$ & 85.44 (75.94, 93.46) & 89.20 (84.98, 93.43) & 93.79 (90.06, 97.13) & 66.67 (51.22, 82.35) & $<0.01$ \\
Planning CT+3D Dose & ViT MM-DiT & PoE & 83.22 (76.32, 89.29) & 89.66 (86.54, 92.55) & 96.62 (94.72, 98.34) & 49.18 (36.92, 61.29) & $<0.01$ & 87.82 (79.35, 95.01) & 90.61 (86.38, 94.37) & 94.92 (91.37, 98.19) & \textbf{69.44} (54.55, 83.34) & $<0.01$ \\
Planning CT+3D Dose & Deterministic MM-DiT & MoE & 84.30 (77.85, 89.87) & 88.94 (85.82, 91.83) & 95.49 (93.19, 97.46) & 50.82 (38.46, 63.23) & - & 87.87 (80.96, 93.97) & 87.79 (83.10, 92.02) & 93.79 (90.05, 97.19) & 58.33 (42.86, 75.01) & - \\
Planning CT+3D Dose & Deterministic MM-DiT & PoE & 85.09 (78.57, 90.65) & 89.90 (86.78, 92.55) & 95.49 (93.31, 97.53) & 57.38 (44.82, 70.37) & - & 88.52 (80.36, 95.28) & 90.14 (86.39, 93.90) & 96.05 (92.97, 98.84) & 61.11 (44.44, 77.14) & - \\
Planning CT+3D Dose & MM-DiT & MoE     & 85.10 (78.46, 90.88) & 89.42 (86.54, 92.31) & 93.80 (91.32, 96.17) & \textbf{63.93} (51.72, 75.81) & $<0.01$ & 88.06 (80.53, 94.41) & 91.08 (86.85, 94.84) & 96.05 (92.86, 98.85) & 66.67 (50.00, 81.40) & $<0.01$ \\
Planning CT+3D Dose & MM-DiT & PoE     & \textbf{85.59} (79.21, 91.27) & \textbf{90.87} (88.22, 93.51) & 96.34 (94.25, 98.08) & 59.02 (46.43, 71.19) & - & \textbf{89.11} (81.49, 95.30) & \textbf{92.02} (88.26, 95.32) & \textbf{97.74} (95.38, 99.45) & 63.89 (47.37, 79.31) & - \\
\bottomrule
\end{tabular}}
\end{table*}

\begin{figure*}[!t]
    \centering
    \begin{subfigure}[t]{0.49\linewidth}
        \centering
        \includegraphics[width=\linewidth]{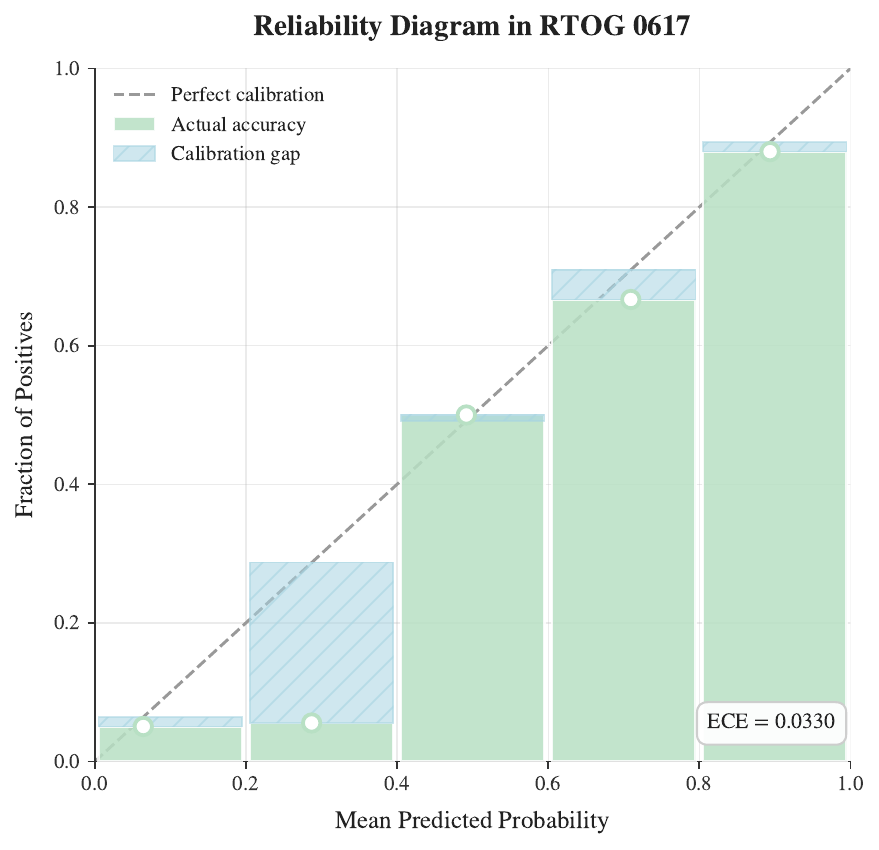}
        \caption{Model calibration and reliability analysis in the RTOG 0617 cohort.}
        \label{fig:calibration_tcia}
    \end{subfigure}
    \hfill
    \begin{subfigure}[t]{0.49\linewidth}
        \centering
        \includegraphics[width=\linewidth]{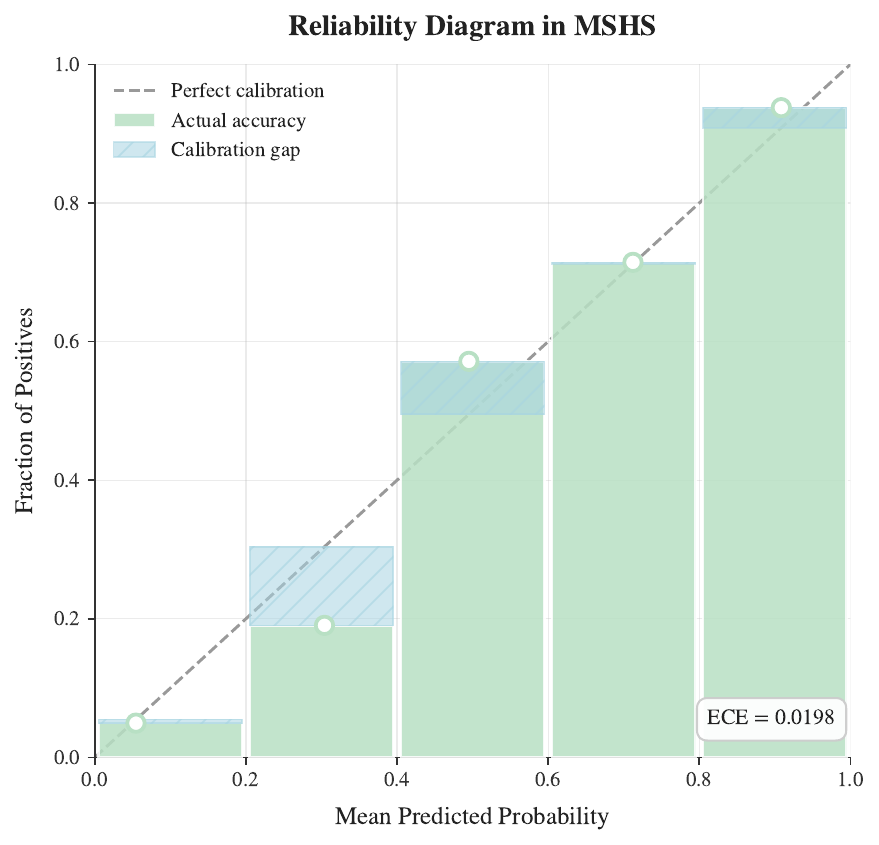}
        \caption{Model calibration and reliability analysis in the MSHS cohort.}
        \label{fig:calibration_ms}
    \end{subfigure}
    \caption{Calibration of RP risk predictions in the (a) RTOG 0617 and (b) MSHS cohorts. The diagonal dashed line represents perfect calibration, and the solid line represents the MM-DiT calibration curve. Expected calibration error (ECE) quantifies the average discrepancy between predicted probabilities and observed RP frequencies. The close agreement between the calibration curves and the ideal calibration line across both cohorts indicates favorable calibration of MM-DiT.}
    \label{fig:calibration}
\end{figure*}

MM-DiT also demonstrated favorable calibration across both cohorts. The model achieved an expected calibration error (ECE) of $0.0330$ in RTOG 0617 and $0.0198$ in MSHS, indicating close agreement between predicted probabilities and observed RP frequencies. Consistent with these results, the calibration curves showed that the predicted risks closely followed the ideal calibration line in both cohorts (Figure \ref{fig:calibration}). These results indicate that MM-DiT provides well-calibrated individualized RP risk estimates in addition to strong discriminative performance.

\subsection{Uncertainty Quantification and Decision-Analytic Utility}
\label{sec22}
\begin{figure*}[!t]
    \centering
    \begin{subfigure}[t]{0.49\linewidth}
        \centering
        \includegraphics[width=\linewidth]{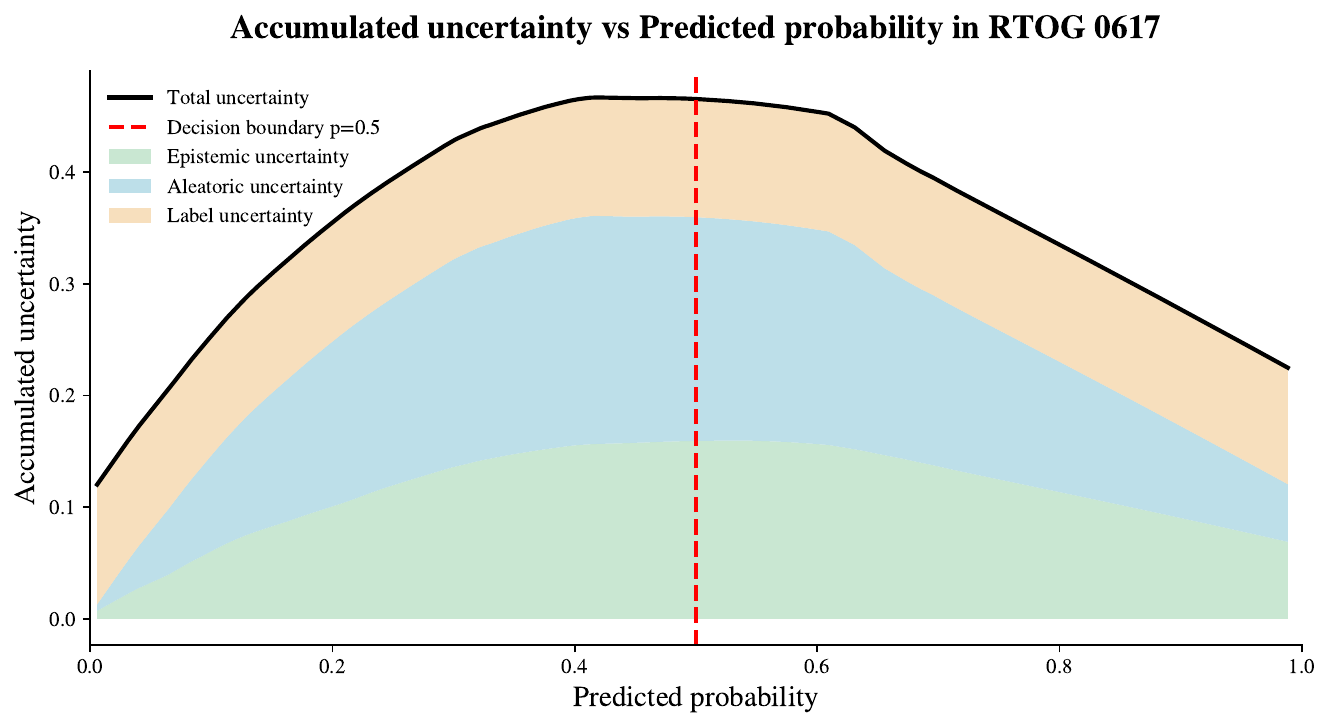}
        \caption{Accumulated predictive uncertainty as a function of predicted RP probability for the proposed MM-DiT Bayesian classifier in the RTOG 0617 cohort.}
        \label{fig:uncertainty_tcia}
    \end{subfigure}
    \hfill
    \begin{subfigure}[t]{0.49\linewidth}
        \centering
        \includegraphics[width=\linewidth]{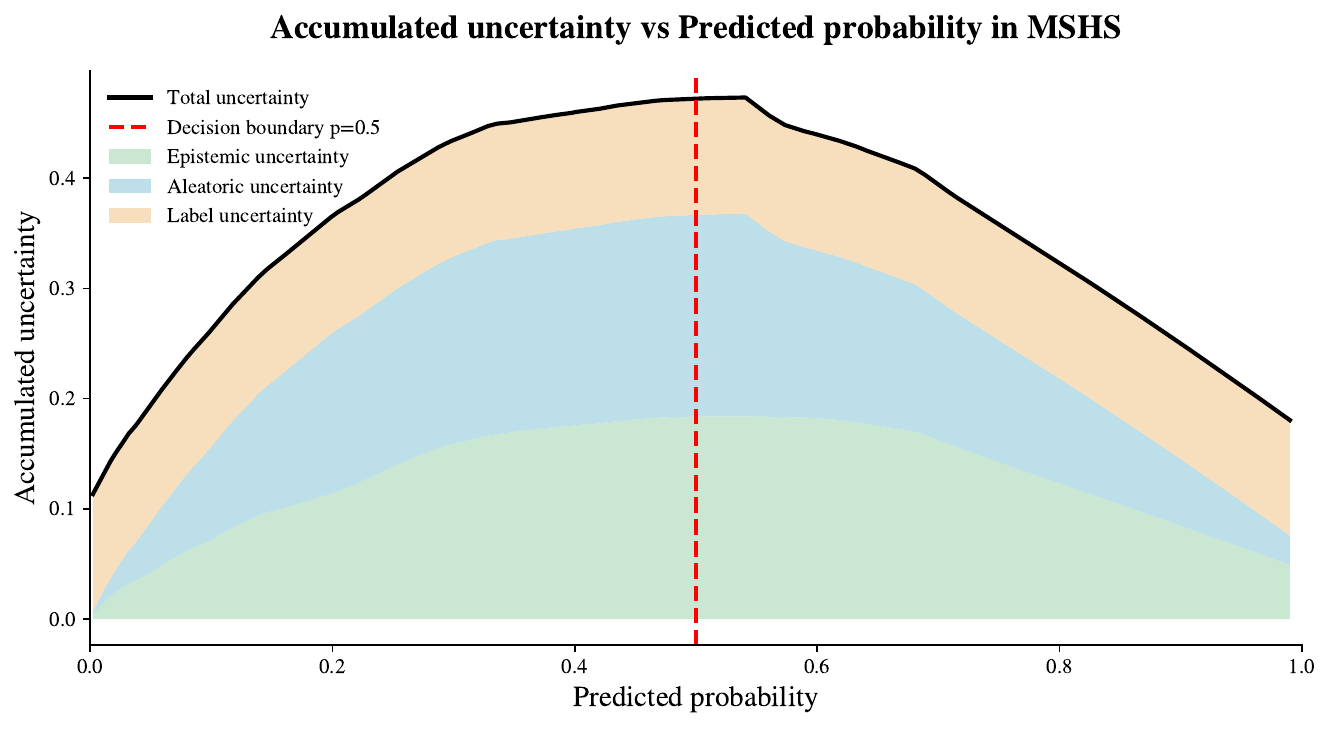}
        \caption{Accumulated predictive uncertainty as a function of predicted RP probability for the proposed MM-DiT Bayesian classifier in the MSHS cohort.}
        \label{fig:uncertainty_ms}
    \end{subfigure}
    \caption{Predictive uncertainty as a function of predicted RP probability across the RTOG 0617 and MSHS cohorts. Total predictive uncertainty is decomposed into aleatoric, epistemic, and label uncertainty. Aleatoric and epistemic uncertainty increase near the decision boundary (predicted probability $\approx 0.5$) and decrease toward the probability extremes, whereas label uncertainty shows a less pronounced increase near the decision boundary.}
    \label{fig:uncertainty}
\end{figure*}

Figure~\ref{fig:uncertainty} shows distributions of predictive uncertainty across RP risk probabilities and its decomposition into aleatoric, epistemic, and label uncertainty in the RTOG 0617 and MSHS cohorts. Aleatoric uncertainty was the predominant contributor to total predictive uncertainty in both cohorts. Both aleatoric and epistemic uncertainty were highest for predictions near the decision boundary (approximately 0.5) and decreased as the predicted RP probability approached 0 or 1. Label uncertainty exhibited a similar but less pronounced pattern, with higher values near the decision boundary. Overall, predictions with intermediate estimated RP risk were associated with greater uncertainty across all three components, indicating reduced confidence in risk estimates for patients whose predicted RP probabilities were closer to the decision boundary.

\begin{figure*}[!t]
    \centering
    \begin{subfigure}[t]{0.49\linewidth}
        \centering
        \includegraphics[width=\linewidth]{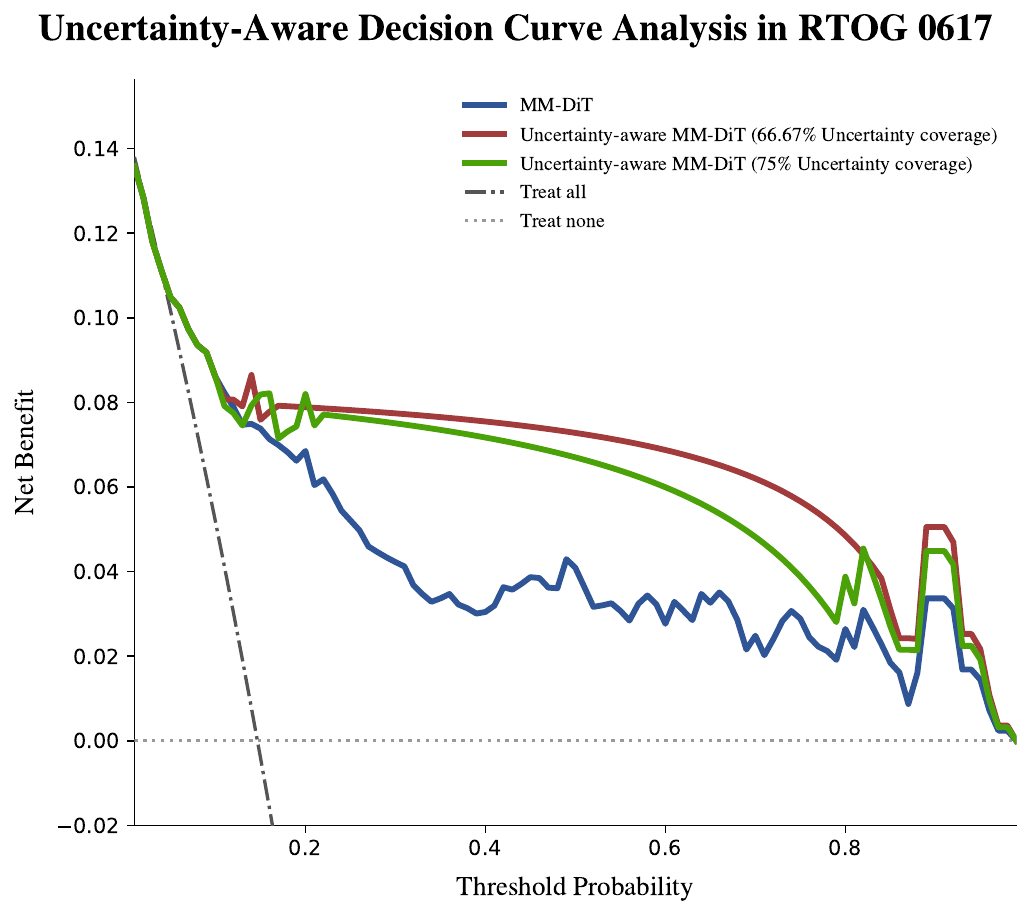}
        \caption{Decision curve analysis in the RTOG 0617 cohort.}
    \end{subfigure}
    \hfill
    \begin{subfigure}[t]{0.49\linewidth}
        \centering
        \includegraphics[width=\linewidth]{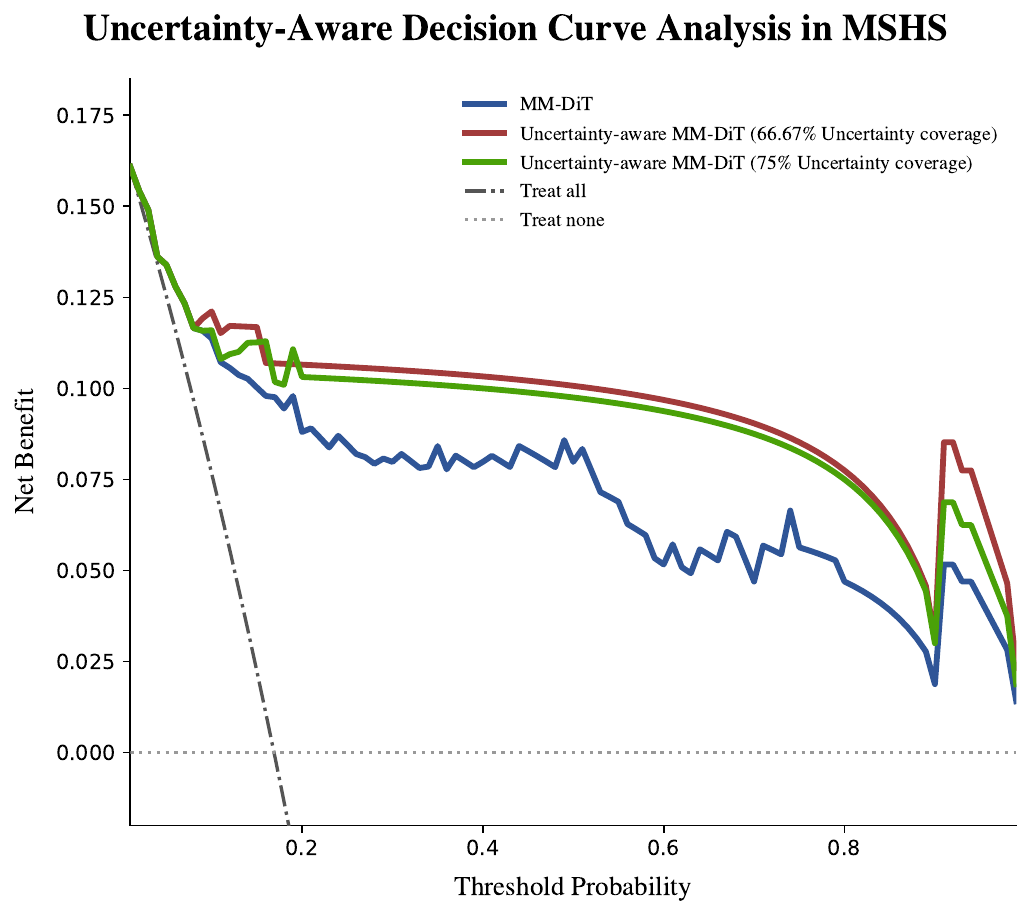}
        \caption{Decision curve analysis in the MSHS cohort.}
    \end{subfigure}
    \caption{Uncertainty-aware decision curve analysis for radiation pneumonitis risk prediction in the RTOG 0617 and MSHS cohorts. Net benefit is shown across a range of threshold probabilities for the conventional MM-DiT model and two uncertainty-aware MM-DiT strategies using the 66.67th- and 75th-percentile total-uncertainty thresholds. Patients with uncertainty above each threshold were excluded from the corresponding selective prediction analysis. “Treat all” and “Treat none” are included as reference decision strategies.}
    \label{fig:decision}
\end{figure*}

To evaluate the clinical utility of uncertainty-aware prediction, DCA was performed and compared with those using selective prediction at different levels of prediction coverage in the RTOG 0617 and MSHS cohorts (Figure \ref{fig:decision}). Specifically, DCA was performed in the whole cohort to examine the potential clinical utility across all patients. Additionally, DCA was further performed on sub-cohorts where patients with high uncertainty were excluded from the selective decision analysis. Two uncertainty thresholds was defined from the empirical distribution of total predictive uncertainty, corresponding to the 66.67th and 75th percentiles, and approximately the highest-uncertainty one-third and one-quarter of patients were excluded, while retaining $66.67\%$ and $75\%$ of the cohort, respectively. Excluding patients with high uncertainty from the selective decision analysis allowed the evaluation of the correlation between the clinical utility and predictive uncertainty, thus highlighting the benefits of quantifying uncertainty in clinical utility. In the RTOG 0617 and MSHS cohorts, net benefit was evaluated across a range of threshold probabilities and compared among the conventional MM-DiT model, the two uncertainty-aware selective prediction strategies, and the standard “treat all” and “treat none” strategies (Figure~\ref{fig:decision}). This analysis evaluated whether selectively withholding predictions for patients with higher predictive uncertainty altered the net benefit of MM-DiT across clinically relevant threshold probabilities. In both cohorts, the MM-DiT curve was above the treat-all and treat-none strategies across the evaluated range of threshold probabilities, suggesting that model-guided decision-making could provide greater net benefit than either default strategy. Furthermore, after excluding patients with the highest predictive uncertainty, the MM-DiT curve showed higher net benefit than the original model among the remaining patients, suggesting that uncertainty-based patient selection may further improve clinical decision-making.

\subsection{Impact of Post-treatment CT on Prediction and Uncertainty}
\label{sec23}
The preceding analyses evaluated MM-DiT using pretreatment information, which does not capture patient-specific changes occurring during and after RT. To examine whether treatment-response information acquired after RT could improve RP prediction, we compared the performance of two MM-DiT configurations in the MSHS cohort: one using planning CT and planned dose alone and another additionally incorporating post-treatment CT. As shown in Table~\ref{tab2}, incorporating post-treatment CT substantially improved prediction performance across all evaluated metrics. The AUC increased from $0.875$ to $0.952$, while accuracy improved from $0.861$ to $0.957$. Specificity and sensitivity also increased from $0.888$ to $0.980$ and from $0.743$ to $0.857$, respectively. These results indicate that post-treatment CT provides complementary information beyond pretreatment anatomy and planned radiation dose for RP prediction.

\begin{table*}[!t]
\centering
\caption{Predictive performance comparison before and after incorporating post-treatment CT in the MSHS cohort using evaluation measures of the area under the receiver operating characteristic curve (AUC), accuracy, specificity, and sensitivity.}
\label{tab2}
\resizebox{\textwidth}{!}{
\begin{tabular}{c|ccccc}
\toprule
Input configuration & AUC ($95\%$CI) & Accuracy ($95\%$CI) & Specificity ($95\%$CI) & Sensitivity ($95\%$CI) & $p$ value \\
\midrule
Planning CT + 3D Dose                     & 87.54 (80.45, 93.78) & 86.10 (81.28, 90.91) & 88.82 (83.56, 93.51)  & 74.29 (58.61, 88.57) & $<0.001$ \\
Planning CT + 3D Dose + Post-treatment CT & 95.15 (89.58, 99.46) & 95.72 (92.51, 98.40) & 98.03 (95.51, 100.00) & 85.71 (72.97, 96.77) & - \\
\bottomrule 
\end{tabular}}
\end{table*}

\begin{figure*}[t]
	\centering
	\includegraphics[width=\linewidth]{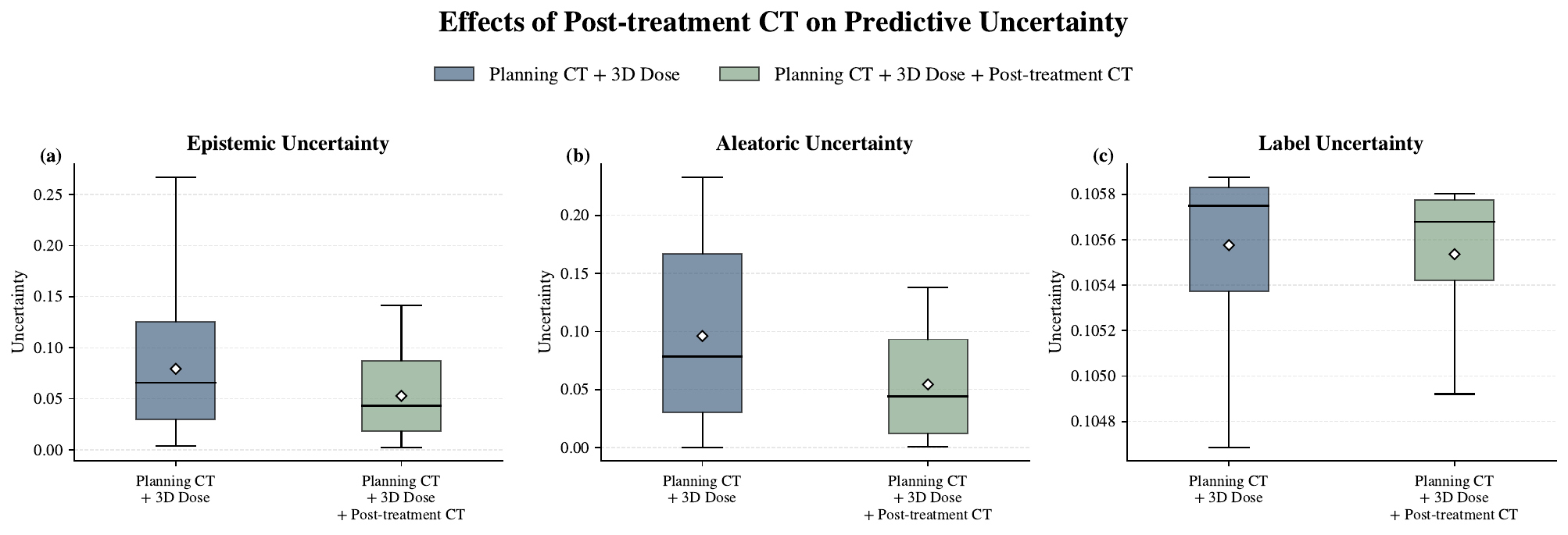} \\
	\caption{Predictive uncertainty with and without post-treatment CT information. Boxplots show the distributions of (a) epistemic uncertainty, (b) aleatoric uncertainty, and (c) label uncertainty for models using pre-treatment CT images and 3D planning dose distributions alone (blue) and models additionally incorporating post-treatment CT images (green).}
	\label{fig:boxplot}
\end{figure*}

To characterize the effect of post-treatment CT information on predictive uncertainty, we compared the distributions of epistemic, aleatoric, and label uncertainty between MM-DiT using planning CT and planned dose alone and that additionally incorporating post-treatment CT (Figure \ref{fig:boxplot}). Incorporating post-treatment CT substantially reduced both epistemic and aleatoric uncertainty. Based on the median uncertainty values, epistemic uncertainty decreased by approximately $40\%$, while aleatoric uncertainty decreased by approximately $40-45\%$. In contrast, label uncertainty changed minimally between the two model configurations, with the median values remaining nearly unchanged. These findings indicate that incorporating post-treatment CT primarily reduced the uncertainty associated with model knowledge and patient-specific response variability, while having limited effect on label uncertainty. 

\begin{figure*}[t]
	\centering
	\includegraphics[width=\linewidth]{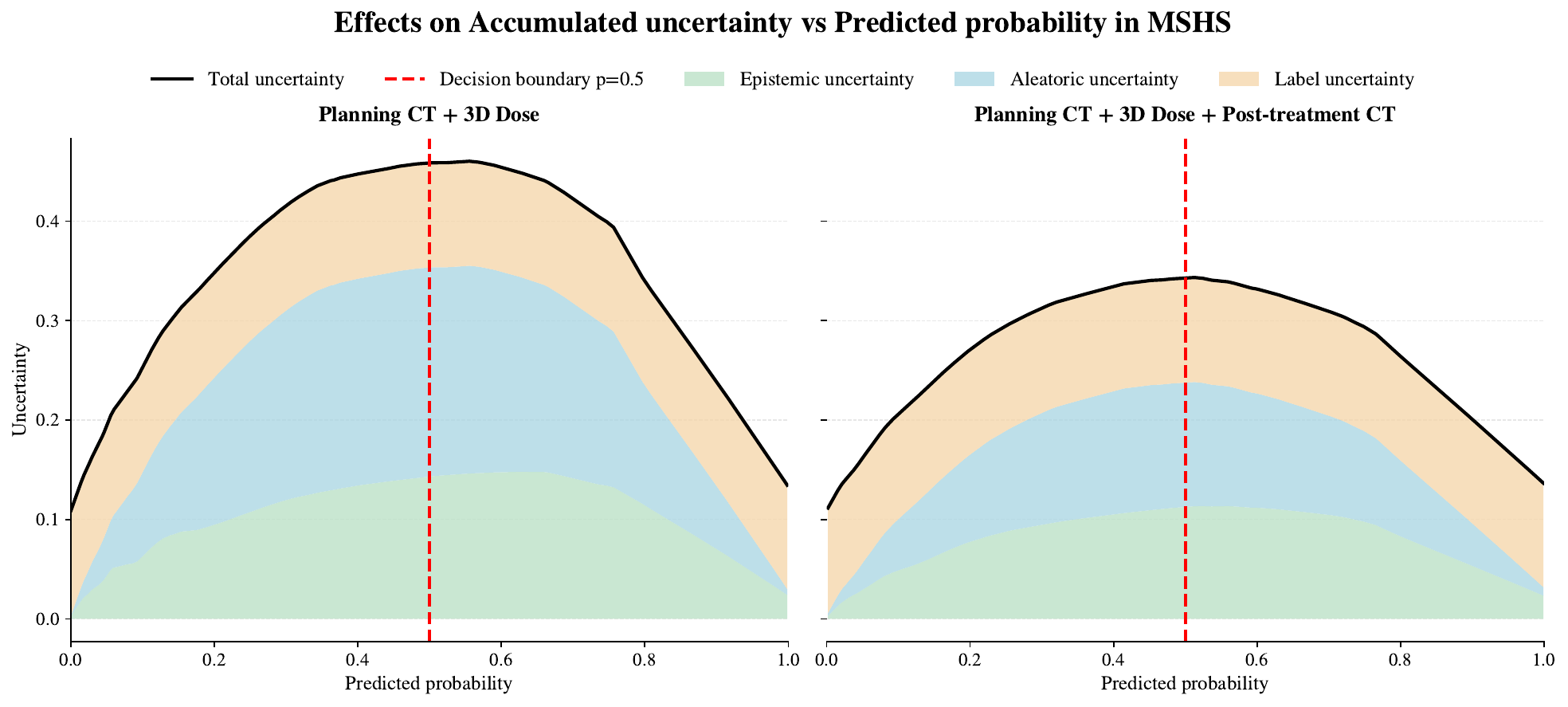} \\
	\caption{Predictive uncertainty across predicted radiation pneumonitis risk. Epistemic, aleatoric, and label uncertainty as a function of predicted probability for models using planning CT images and dose alone (left) and additionally with (right) post-treatment CT.}
	\label{fig:accu_un}
\end{figure*}

To further characterize how post-treatment CT information affects predictive uncertainty across the RP risk spectrum, we examined the distributions of epistemic, aleatoric, and label uncertainty as a function of predicted RP probability (Figure \ref{fig:accu_un}). Compared with the model using planning CT and dose alone, incorporation of post-treatment CT information resulted in lower uncertainty across the predicted-risk range. Reductions were observed across multiple uncertainty components, indicating that the additional treatment-response information altered the uncertainty associated with individual risk estimates, for whom incorporation of post-treatment CT resulted in more concentrated uncertainty distributions across the predicted-risk spectrum.

\subsection{External Validation of Model Generalizability}
\label{sec24}
To assess model generalizability, models trained on the RTOG 0617 cohort were directly evaluated in the independent MSHS cohort without additional adaptation or fine-tuning. As shown in Table~\ref{tab3}, MM-DiT with the PoE fusion achieved the highest discriminative performance, with an AUC of $0.828$. This performance exceeded that of the other evaluated multimodal configurations, including VAE-based models. Within the MM-DiT framework, PoE fusion also outperformed MoE fusion, and CNN-based models outperformed ViT-based models. These findings indicate that MM-DiT with PoE fusion retained strong discriminative performance when applied to an independent cohort, providing evidence of model generalizability across cohorts.

\begin{table*}[!t]
\centering
\caption{External evaluation on multimodal models for radiation pneumonitis prediction using pretreatment planning CT images and radiation dose distributions. Models trained on the RTOG 0617 cohort were directly evaluated in the independent MSHS cohort using the area under the receiver operating characteristic curve (AUC), accuracy, specificity, and sensitivity. \textbf{Bold} represents the best performance. $p_{\mathrm{adj}}$ denotes the false discovery rate-adjusted $p$-value.}
\label{tab3}
\resizebox{\textwidth}{!}{
\begin{tabular}{c|c|ccccc}
\toprule
\multirow{2}{*}{Models} & \multirow{2}{*}{Fusion} &  \multicolumn{5}{c}{RTOG 0617 $\rightarrow$ MSHS} \\
\cmidrule(lr){3-7}
& & AUC ($95\%$CI) & Accuracy ($95\%$CI) & Specificity ($95\%$CI) & Sensitivity ($95\%$CI) & $p_{\tiny adj}$ \\
\midrule
VAE & Concatenation & 69.70 (58.89, 79.33) & 81.69 (76.51, 86.85) & 92.09 (87.84, 95.79) & 30.56 (15.38, 46.88) & $<0.001$ \\
VAE & MoE           & 73.51 (63.37, 82.99) & 84.04 (78.87, 88.73) & 91.53 (87.22, 95.51) & 47.22 (31.03, 63.64) & $<0.001$ \\
VAE & PoE           & 74.86 (63.95, 84.23) & 84.98 (80.28, 89.67) & \textbf{93.22} (89.50, 96.69) & 44.45 (29.03, 60.87) & $<0.001$ \\
MM-DiT-ViT & MoE    & 77.02 (64.64, 87.43) & 85.10 (79.93, 89.79) & 88.14 (82.94, 92.86) & 58.33 (41.38, 73.53) & $<0.001$ \\
MM-DiT-ViT & PoE    & 78.00 (67.58, 86.79) & 85.41 (80.34, 90.20) & 89.27 (84.61, 93.49) & 61.11 (44.11, 77.15) & $<0.001$ \\
MM-DiT & MoE        & 80.86 (70.43, 90.14) & 85.45 (80.75, 90.14) & 91.53 (87.35, 95.43) & 55.56 (39.27, 71.88) & $<0.01$ \\
MM-DiT & PoE        & \textbf{82.80} (71.44, 92.32) & \textbf{87.32} (83.10, 91.55) & 91.53 (87.36, 95.11) & \textbf{66.67} (50.00, 82.05) & - \\
\bottomrule
\end{tabular}}
\end{table*}

\begin{figure*}[t]
	\centering
	\includegraphics[width=\linewidth]{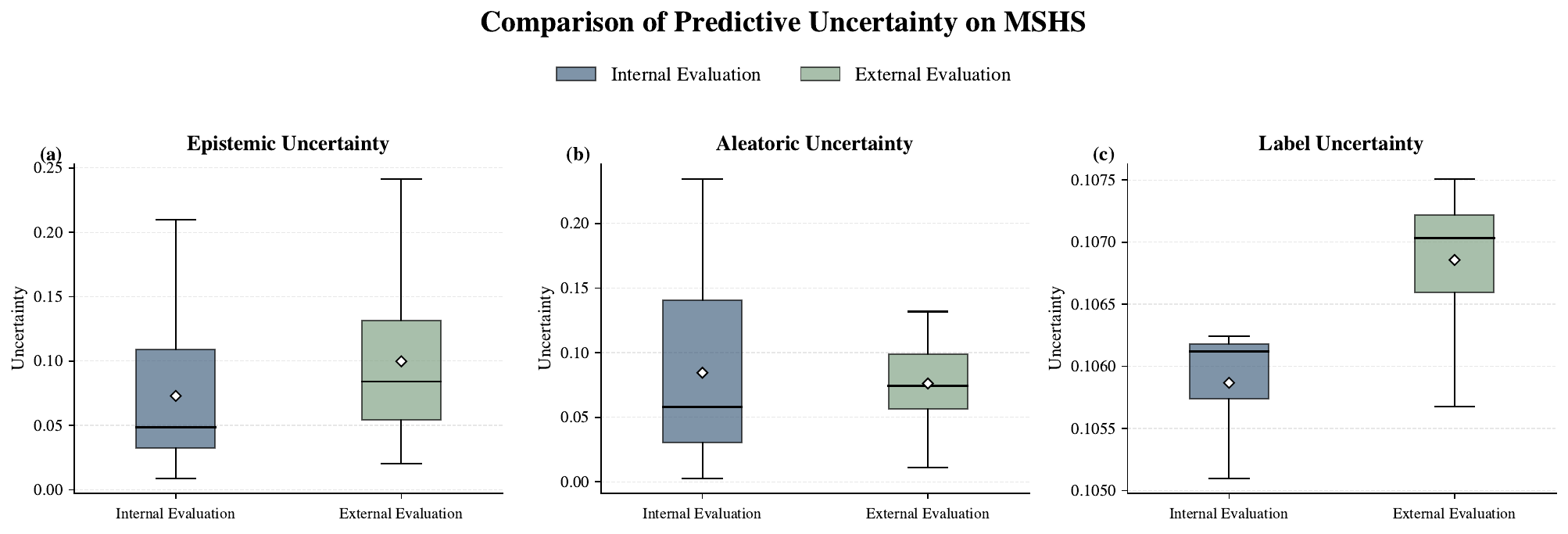} \\
	\caption{Predictive uncertainty of internal (RTOG 0617) and external (MSHS) evaluation. Boxplots show the distributions of (a) epistemic uncertainty, (b) aleatoric uncertainty, and (c) label uncertainty for models using planning CT and dose data of internal evaluation (blue) and external evaluation (green).}
	\label{fig:ex_boxplot}
\end{figure*}

Beyond discrimination, we examined whether predictive uncertainty remained stable under external evaluation (Figure~\ref{fig:ex_boxplot}). The distributions of epistemic, aleatoric, and label uncertainty were broadly comparable between the internal (RTOG 0617) and external (MSHS) evaluations, with no substantial increase in any uncertainty component after application of the model to the independent cohort. This consistency suggests that the uncertainty estimates of MM-DiT with PoE fusion were largely preserved under external evaluation, complementing its discriminative performance and providing additional evidence of generalizability across cohorts.

\subsection{Illustrative Cases of Uncertainty-Aware RP Prediction}
\label{sec25}
%For individual patients, the proposed framework provides a case-level interpretation that jointly considers the predicted radiation pneumonitis (RP) risk and its associated uncertainty, rather than relying on predefined thresholds to categorize uncertainty. For each patient, the model provides an estimated probability of RP, the corresponding predicted RP outcome, and three complementary components of uncertainty: epistemic, aleatoric, and label uncertainty. Specifically, the model output can be summarized as

%\begin{itemize}
%\item Predicted RP risk: $\hat{p}$
%\item Predicted RP grade: $\hat{y}$
%\item Aleatoric (data) uncertainty: $\pm U_{\mathrm{alea}}$
%\item Epistemic (model) uncertainty: $\pm U_{\mathrm{epst}}$
%\item Label uncertainty: $\pm U_{\mathrm{label}}$
%\end{itemize}

%The predicted probability represents the model's estimated likelihood of RP, whereas the decomposed uncertainty measures provide complementary information about the confidence and limitations associated with this estimate. Aleatoric uncertainty reflects ambiguity or irreducible variability inherent in the available patient data, epistemic uncertainty reflects uncertainty in the model's learned representation and its ability to generalize, and label uncertainty captures potential ambiguity or inconsistency in the reference outcome. These components are not interpreted as independent measures of model performance; rather, they are considered jointly with the predicted risk to characterize the reliability and informativeness of an individual prediction.

We present three illustrative cases demonstrating uncertainty-aware RP risk prediction and the impact of incorporating post-treatment CT (Table~\ref{tab4}). The cases represent three complementary scenarios: (1) a pretreatment prediction near the decision boundary with high predictive uncertainty; (2) substantial risk reclassification after incorporating post-treatment CT; and (3) a marked reduction in predictive uncertainty following incorporation of post-treatment CT.

\noindent\textbf{Case 1: Boundary prediction with substantial pretreatment uncertainty.}
Using planning CT and 3D dose alone, MM-DiT estimated an RP risk of 0.521, with relatively high epistemic ($U_{\mathrm{epst}}=0.152$) and aleatoric ($U_{\mathrm{alea}}=0.226$) uncertainty. After incorporating post-treatment CT, the estimated risk decreased to 0.014, while epistemic and aleatoric uncertainty decreased to 0.061 and 0.010, respectively; label uncertainty remained essentially unchanged. Thus, the borderline pretreatment prediction shifted toward a low-risk estimate with substantially lower predictive uncertainty.

\noindent\textbf{Case 2: Substantial risk reclassification after post-treatment CT.}
The pretreatment model estimated an RP risk of 0.141, which increased to 0.938 after incorporating post-treatment CT. Despite this marked change, epistemic and aleatoric uncertainty decreased from 0.117 to 0.085 and from 0.108 to 0.051, respectively, while label uncertainty remained nearly unchanged. Thus, treatment-response imaging shifted the prediction from low to high risk while simultaneously reducing predictive uncertainty.

\noindent\textbf{Case 3: Large reduction in predictive uncertainty.} 
The pretreatment model estimated an RP risk of 0.482, with epistemic and aleatoric uncertainties of 0.142 and 0.229, respectively. After incorporating post-treatment CT, the estimated risk increased to 0.961, while epistemic uncertainty decreased to 0.067 and aleatoric uncertainty to 0.033; label uncertainty remained essentially unchanged (0.105 versus 0.104). The predictive results on this patient showed a large reduction in the combined epistemic and aleatoric uncertainty, demonstrating that treatment-response imaging can substantially reduce uncertainty while refining the estimated RP risk.

\begin{figure*}[t]
	\centering
	\includegraphics[width=0.9\linewidth]{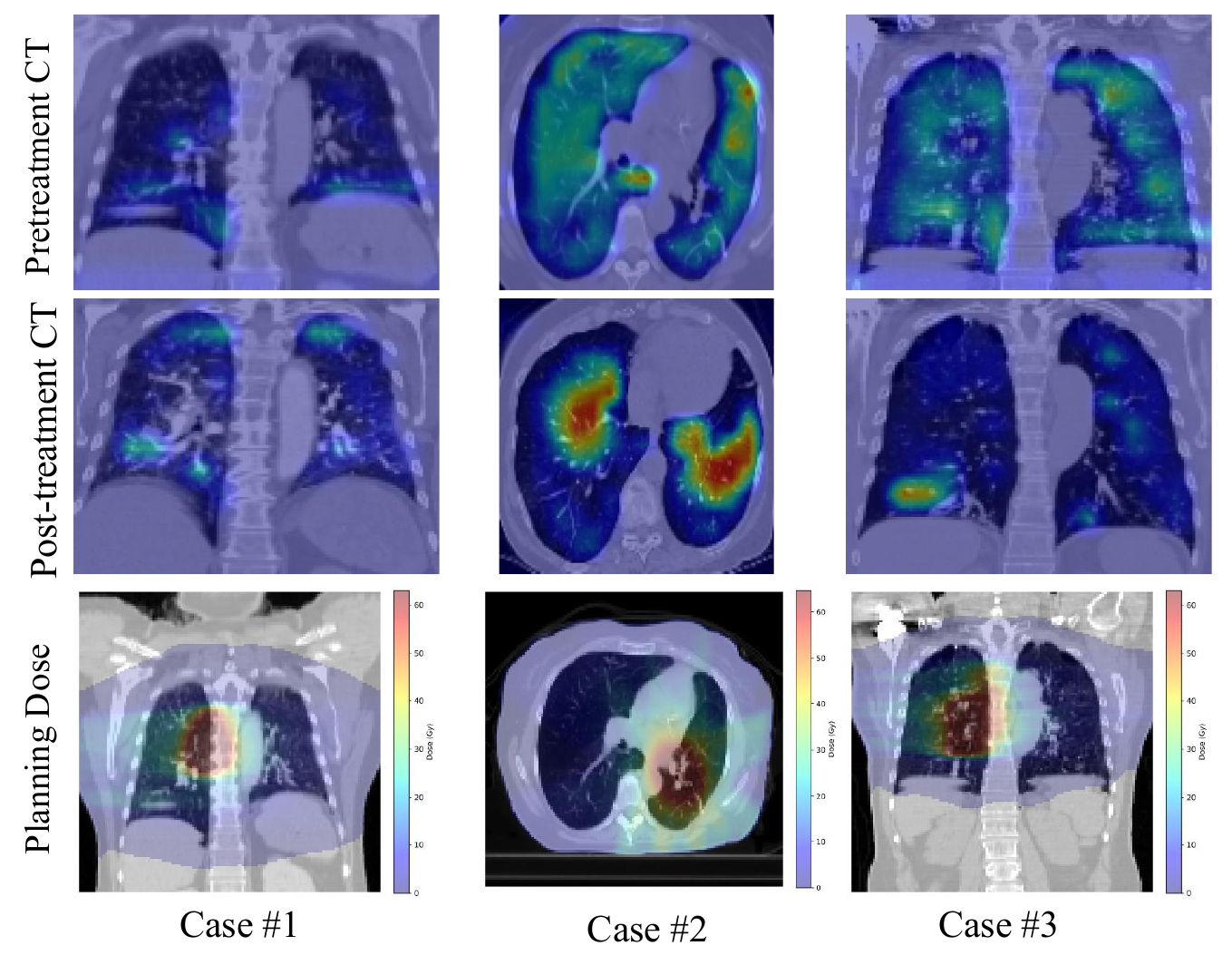} \\
	\caption{Grad-CAM visualization of anatomical regions contributing to radiation pneumonitis (RP) predictions and their spatial relationship with radiation dose. Three representative cases are shown (Cases \#1-3) from the MSHS cohort are shown. The top and middle rows show Grad-CAM overlays on pretreatment and post-treatment CT images respectively, illustrating changes in model attribution following incorporation of treatment-response imaging. The bottom row shows the corresponding planning CT images with radiation dose distributions. Warmer colors in the Grad-CAM maps indicate greater relative model attribution, whereas the dose overlays represent the planned radiation dose (Gy). The spatial correspondence and differences between model-attributed regions and irradiated regions illustrate that regions contributing to RP predictions do not necessarily coincide with the highest-dose regions. Grad-CAM reflect model attribution rather than direct measurement of radiation dose or radiation-induced injury.}
	\label{fig:interpretation}
\end{figure*}

To further examine model behavior, we used 3D Grad-CAM to visualize anatomical regions contributing to predictions in three representative cases (Figure~\ref{fig:interpretation}). In the planning CTs, highlighted regions showed spatial correspondences with or were adjacent to pulmonary structures receiving substantial radiation dose. Incorporation of post-treatment CT revealed additional or altered activation patterns in regions exhibiting post-treatment imaging changes, particularly in cases with substantial changes in predicted risk. These visualizations illustrate how MM-DiT uses pretreatment anatomical information and post-treatment imaging features to update individualized RP risk estimates, while highlighting that model-attributed regions do not necessarily coincide with the highest-dose regions. Importantly, Grad-CAM identifies regions associated with model predictions and does not establish causal mechanisms or directly identify radiation-induced injury. Together, these case studies illustrate the potential of MM-DiT to integrate longitudinal imaging information for updating individualized RP risk and predictive uncertainty as treatment-response information becomes available.

\begin{table*}[!t]
\centering
\caption{Three illustrative patient cases showing changes in predicted radiation pneumonitis risk and decomposed predictive uncertainty before and after incorporating post-treatment CT. Uncertainty is decomposed into epistemic, aleatoric, and label components.}
\label{tab4}
\resizebox{\textwidth}{!}{
\begin{tabular}{c|c|cccc|cccc}
\toprule
\multirow{2}{*}{Case} & Diagnosis & \multicolumn{4}{c|}{Before treatment (Planning CT + 3D Dose)} & \multicolumn{4}{c}{After treatment (Planning CT + 3D Dose + Post-treatment CT)} \\
\cmidrule(lr){2-10}
& RP label & RP risk & $U_{\mathrm{epst}}$ & $U_{\mathrm{alea}}$ & $U_{\mathrm{label}}$ & RP risk & $U_{\mathrm{epst}}$ & $U_{\mathrm{alea}}$  & $U_{\mathrm{label}}$ \\
\midrule
1 & 0 & 0.521 & 0.152 & 0.226 & 0.105 & 0.014 & 0.061 & 0.010 & 0.105 \\
2 & 1 & 0.141 & 0.117 & 0.108 & 0.106 & 0.938 & 0.085 & 0.051 & 0.105 \\
3 & 1 & 0.482 & 0.142 & 0.229 & 0.105 & 0.961 & 0.067 & 0.033 & 0.104 \\
\bottomrule
\end{tabular}}
\end{table*}

\section{Discussion}
\label{sec3}
MM-DiT demonstrated consistent predictive performance across two independent cohorts, outperforming unimodal and alternative multimodal models in discrimination and overall classification performance (Table~\ref{tab1} and Figure \ref{fig:roc}). Importantly, the model also showed favorable calibration, with ECEs of 0.033 and 0.020 in the RTOG 0617 and MSHS cohorts, respectively (Figure \ref{fig:calibration}). Thus, the model provided not only strong discrimination but also probability estimates that closely corresponded to observed RP frequencies, an important distinction for clinical risk assessment where absolute risk estimates are more informative than relative ranking alone. The preservation of discrimination and calibration across independent cohorts provides supporting evidence for the robustness of the learned multimodal representations. 

Beyond predictive performance, uncertainty quantification provided information about the reliability of individual risk estimates. Aleatoric uncertainty was the dominant component and was greatest near the decision boundary (Figure~\ref{fig:uncertainty}), suggesting that substantial uncertainty may reflect patient-specific variability in response to radiation that is not fully captured by pretreatment information. Such uncertainty is unlikely to be substantially reduced through model complexity or training-set size alone and may instead require additional information about the patient's evolving response, including longitudinal clinical assessments and imaging  \cite{xiao2018comparison,du2020novel,ajdari2022personalized,azumi2023early,bucknell2023mid,klaar2024mri}. Epistemic uncertainty provided  complementary information about limitations in model knowledge, including potential differences between patient populations, imaging characteristics, and treatment patterns across cohorts. Its characterization may therefore help identify predictions for which the available training data provide limited support, an important consideration for deployment across institutions \cite{faghani2023quantifying}. Label uncertainty captured a distinct source of uncertainty arising from ambiguity or inconsistency in observed RP outcomes, which is particularly relevant for RP given potential variation in grading, documentation, timing, or interpretation \cite{faria2009challenge,tucker2010impact,fairchild2020interrater}. Together, these components provide information about the reliability of an individual prediction that cannot be conveyed by the predicted probability alone.

The incorporation of post-treatment CT further demonstrated how longitudinal information can improve both risk estimation and prediction confidence (Table~\ref{tab2} and Figure \ref{fig:boxplot}). Adding post-treatment CT substantially improved discrimination while reducing aleatoric and epistemic uncertainty by approximately $40\%$, with minimal change in label uncertainty (Figure \ref{fig:boxplot}). These findings suggest that treatment-response imaging provides patient-specific information beyond pretreatment anatomy and planned dose, enabling the model to refine risk estimates as new evidence becomes available. The greater reduction in uncertainty among patients with higher pretreatment uncertainty further indicates that longitudinal imaging may be particularly informative when the initial prediction is less certain. Notably, the illustrative cases showed distinct patterns of this longitudinal updating (Table~\ref{tab4}): post-treatment CT resolved a borderline, high-uncertainty prediction in one patient, substantially reclassified risk while reducing uncertainty in another, and produced a pronounced reduction in uncertainty in a third. these cases emphasize that changes in predicted risk should be interpreted together with changes in predictive uncertainty and illustrate how additional treatment-response information can refine patient-specific risk assessment. The accompanying 3D Grad-CAM maps provide a spatial perspective on how the model integrates imaging information for RP prediction (Figure~\ref{fig:interpretation}). Although some model-attributed regions spatially corresponded to irradiated lung regions, the observed activation patterns were not consistently confined to the highest dose regions. This suggests that the model may incorporate anatomical and imaging characteristics that are not fully represented by conventional dose distributions. Differences between pretreatment and post-treatment activation patterns further illustrate the potential value of longitudinal imaging for characterizing evolving pulmonary changes. Nevertheless, these visualizations should be interpreted as model attribution rather than direct evidence of radiation-induced injury, and the biological significance of the highlighted regions requires further investigation.

Collectively, these findings support a longitudinal framework for uncertainty-aware RP prediction in which risk assessment evolves from an initial pretreatment estimate to progressively informed patient-specific predictions as new evidence becomes available. Rather than treating the pretreatment prediction as a fixed probability, such a framework can characterize both the estimated risk and the degree of uncertainty surrounding it, while allowing subsequent imaging and clinical information to update both quantities. Predictions with greater uncertainty could consequently be prioritized for additional clinical review or reassessment, whereas lower-uncertainty predictions may provide stronger support for risk stratification. Importantly, the representative cases are intended to illustrate model behavior rather than establish clinical thresholds for obtaining additional imaging, intensified surveillance, or treatment modification. Prospective validation will be necessary to determine whether longitudinal, uncertainty-guided risk updating and model interpretation can translate into actionable improvements in clinical decision-making and patient outcomes.

The potential clinical utility of MM-DiT was further evaluated using the decision curve analysis (Figure~\ref{fig:decision}), which showed positive net benefit across a broad range of threshold probabilities compared with treat-all and treat-none strategies \cite{vickers2021decision,zhang2023radiomics}. These findings suggest that MM-DiT may provide information relevant to clinical risk stratification beyond discrimination and calibration, with potential benefit across a range of decision thresholds. Importantly, the risk–uncertainty framework is not intended to prescribe treatment modifications based solely on model output. Rather, it provides a mechanism for interpreting predicted RP risk together with its associated uncertainty, allowing higher-uncertainty predictions to be identified for additional clinical assessment. Prospective studies will be needed to determine whether incorporating predictive uncertainty into RP risk assessment can translate into improved toxicity monitoring, treatment planning, or other aspects of clinical management.

Several limitations should be considered. First, this is a retrospective study, and the primary model was developed using pretreatment CT and radiation dose distributions without incorporating clinical variables, such as demographic characteristics, smoking history, baseline pulmonary function, or comorbidities. This decision was partly driven by the limited availability and inconsistent documentation of these variables across the MSHS and RTOG 0617 cohorts, particularly in RTOG 0617, which restricted the selection of harmonized clinical inputs for model development and external evaluation.  
Although imaging–dosimetric inputs provide complementary anatomical and treatment-related information, they do not capture the full spectrum of patient-specific factors associated with RP susceptibility or the evolving biological response to RT. The omission of clinical variables may therefore limit individualized risk prediction, particularly among patients with similar anatomical and dosimetric characteristics but different baseline risk profiles. Future studies should investigate whether integrating harmonized clinical variables improves predictive performance, calibration, and uncertainty characterization. In addition, our analysis incorporating post-treatment CT provides preliminary evidence that longitudinal imaging can improve risk estimation and reduce predictive uncertainty; however, it does not yet represent a prospective model that continuously updates predictions using weekly on-treatment information. Future studies should therefore develop and prospectively evaluate longitudinal models that sequentially integrate treatment-response imaging, delivered dose, clinical assessments, and patient-reported symptoms throughout and after RT. The predominance of aleatoric uncertainty observed in our analysis further supports investigating whether richer longitudinal data can better characterize patient-specific variability in radiation response.

Second, although MM-DiT was evaluated in two independent cohorts and demonstrated preserved predictive performance under external evaluation, broader multi-institutional validation is needed to assess generalizability across diverse patient populations, imaging protocols, treatment techniques, and clinical practice patterns. Such studies should also evaluate whether the individual uncertainty estimates remain well calibrated when the model is applied to populations that differ from the development cohorts. Third, RP outcomes may be subject to variability in clinical grading, documentation, timing, and interpretation. Although the label-noise component of MM-DiT explicitly accounts for uncertainty in outcome labels, residual uncertainty in toxicity assessment cannot be eliminated through modeling alone. Prospective studies using standardized toxicity assessment and harmonized longitudinal follow-up may help reduce this source of uncertainty and provide more reliable reference outcomes. Fourth, the risk and uncertainty thresholds used for the four-group risk–uncertainty stratification were selected for analytical illustration and should not be interpreted as clinical decision thresholds. Their clinical relevance and optimal operating points will require prospective evaluation in specific clinical workflows and patient populations. Finally, the potential clinical utility suggested by DCA does not establish prospective clinical effectiveness. The model has not yet been evaluated for its ability to change clinician behavior, improve toxicity monitoring, alter treatment decisions, or improve patient outcomes. Future prospective studies should therefore assess how uncertainty-aware longitudinal predictions can be integrated into real-world thoracic radiation oncology workflows and whether such integration improves clinical decision-making and patient-centered outcomes.

%H3: Incorporation of post-treatment CT provides incremental patient-specific information that improves discrimination, updates risk estimates, and reduces predictive uncertainty, particularly among patients with high pretreatment uncertainty. The value of additional imaging is greatest when the initial model is uncertain (1) Incremental performance by large information gain: longitudinal updating: improve AUC, improve calibration, reduce uncertainty, enhance clinical utility.

\section{Methods}
\label{sec4}
\subsection{Patient cohorts}
The proposed framework is developed and evaluated using data from two independent patient cohorts. Patients were categorized into two groups according to RP severity: no-RP2 (RP grade $\leq$ 1) and RP2 (RP grade $\geq$ 2). Each patient record includes a planning CT image set and corresponding 3D spatial dose distribution. The first cohort is derived from the NRG Radiation Therapy Oncology Group (RTOG) 0617 randomized clinical trial and accessed through The Cancer Imaging Archive \cite{Bradley2018NSCLC} with approval from the National Cancer Institute (NCI) National Clinical Trials Network (NCTN)/National Community Oncology Research Program (NCORP). The trial includes 544 patients with stage III non-small-cell lung cancer (NSCLC) who were enrolled in a randomized study evaluating radiation dose escalation in combination of chemotherapy, with and without cetuximab \cite{bradley2015standard,movsas2016quality}. Of the 544 patients, 128 are excluded because of missing CT or dose data ($n = 56$), missing RP grade ($n = 31$), or mismatched dimensions among CT images, dose distributions, and lung masks ($n = 41$). A total of 416 subjects are therefore included from the RTOG 0617 cohort. The second cohort comprised 213 patients with lung cancer who received thoracic RT at the Mount Sinai Health System (MSHS). 187 patients from the MSHS cohort had post-treatment CTs for diagnosis of PR within six months after the radiation treatment. For each patient, the planning CT image set, segmented lung masks, and corresponding 3D dose distributions are retrieved and de-identified for model development and validation. Detailed demographic and clinical characteristics for these two cohorts are summarized in Table \ref{tab_demo}.

\begin{table*}[!t]
	\centering
	\caption{Comparison of the demographic, clinical and tumor biology characteristics of two cohorts. Mean and [Min, Max] for age are shown. Absolute and relative frequencies $(\%)$ for categorical variables are shown.}
	\label{tab_demo}
	\resizebox{0.9\textwidth}{!}{
		\begin{tabular}{c  c | c | c}
			\toprule
			\multicolumn{2}{c|}{Cohorts} & RTOG 0617 & MSHS \\
			\hline
			\multicolumn{2}{c|}{Population Number}    & 416 & 213 \\
			\hline
			\multicolumn{2}{c|}{Age} &   63.23$\pm$9.12 [37, 83] & 68.03$\pm$10.75 [36, 90] \\
			\hline
			\multirow{2}{*}{Gender} &
			Female & 171 ($41.11\%$) & 101 ($47.42\%$) \\
			&	Male   & 245 ($58.89\%$) & 112 ($52.58\%$) \\
			\hline
			\multirow{5}{*}{Histology} &
			Adenocarcinoma              & 164 ($39.42\%$) & 96 ($45.07\%$) \\
			&	Large cell undifferentiated & 12  ($2.88\%$)  & 22 ($10.33\%$) \\
			&	Other                       & 62  ($14.90\%$) & 6  ($2.82\%$) \\
			&	Squamous cell carcinoma     & 178 ($42.79\%$) & 81 ($38.03\%$) \\
			&	Unknown                     & 0   ($0.00\%$)  & 8  ($3.76\%$) \\
			\hline
			\multirow{7}{*}{Stage} &
			Ia      & 0   ($0.00\%$)  & 0 ($0.00\%$) \\
			&	Ib      & 0   ($0.00\%$)  & 0 ($0.00\%$) \\
			&	IIa     & 0   ($0.00\%$)  & 21 ($9.86\%$) \\
			&	IIb     & 0   ($0.00\%$)  & 14 ($6.57\%$) \\
			&	IIIa    & 278 ($66.83\%$) & 96 ($45.07\%$) \\
			&	IIIb    & 138 ($33.17\%$) & 59 ($27.70\%$) \\
			&	IV      & 0   ($0.00\%$)  & 13 ($6.10\%$) \\
			&	Unknown & 0   ($0.00\%$)  & 10 ($4.69\%$) \\
			\hline
			\multirow{2}{*}{Technique} &
			3D CRT & 230 ($55.29\%$) & 49 ($23.00\%$) \\
			&	IMRT   & 186 ($44.71\%$) & 164 ($77.00\%$) \\
			\hline
			\multirow{4}{*}{Smoking} &
			Current Smoker & 197 ($47.36\%$) & 34  ($15.96\%$) \\
			& Former Smoker  & 172 ($41.35\%$) & 158 ($74.18\%$) \\
			& Non-Smoker     & 27  ($6.49\%$)  & 18  ($8.45\%$) \\
			& Unknown        & 20  ($4.81\%$)  & 3   ($1.41\%$) \\
			\bottomrule
	\end{tabular}}
\end{table*}

\subsection{Data Preprocessing}
The pre-treatment CT images and 3D planning dose distributions undergo a multi-step preprocessing pipeline prior to model training. First, to mitigate the influence of outliers and noise, voxel intensities are clipped using window ranges of $(-1000, 400)$ HU for CT images and $(0, 90)$ Gy for dose distributions. Following intensity clipping, both modalities are cropped to the thoracic foreground using mask-guided cropping based on the respective lung masks. These cropped volumes are then resampled to a fixed isotropic dimension of $128\times128\times96$ voxels. Finally, $z$-score normalization is applied to the rescaled volumes to ensure numerical stability and robust feature learning during the training process. To ensure the robust training, data augmentation techniques were employed. Specifically, preprocessed CT and dose patches were mirrored along two axes ($x$ and $y$) with a probability of $0.5$. CT and dose patches were also transformed via affine transformation with a probability of $0.3$. Zero-centered additive Gaussian noise with variance drawn from the distribution $U(0, 0.01)$, and brightness adjustments were added to each CT sample voxel with a probability of $0.2$ and $0.3$, respectively.

Five-fold cross-validation was performed independently within each cohort. In each fold, approximately $80\%$ of patients were used for model training and the remaining $20\%$ were held out for testing, such that each patient was included in the test set once. A fixed random state of 42 was used for reproducible data partitioning. The same preprocessing procedures, self-supervised pre-training protocols, model initialization strategies, and fine-tuning configurations were applied consistently across MM-DiT and all compared models to ensure a fair and consistent experimental setting.

\subsection{Model Development and Implementation}
The overall framework is illustrated in Figure \ref{fig:model}. MM-DiT is a multimodal deep learning framework designed for RP risk prediction and uncertainty quantification through two sequential stages: \textbf{self-supervised multimodal representation learning} and \textbf{supervised Bayesian fine-tuning}. 

\textbf{Self-supervised Multimodal Representation Learning.} During self-supervised pre-training, a multi-modal Diffusion Transformer learns cross-modal latent representations extracted from pre-treatment CT images and corresponding 3D planning dose distributions without using RP labels. Specifically, MM-DiT takes CT images and corresponding 3D dose distributions as two input modalities. Modality-specific encoders with an identical architecture are used to extract features from each modality. Both convolutional neural network (CNN)-based and Vision Transformer (ViT)-based encoders are evaluated to assess the effect of encoder architecture on multimodal representation learning. For the CNN-based encoder, the network consists of three stages, with each stage comprising a 3D $3\times3\times3$ convolutional layer with a stride of 2 followed by a ReLU activation. The spatial dimensions of the feature maps are therefore downsampled by a factor of 2 at each stage, while the number of feature channels increases to 32, 64, and 128, respectively. The ViT-based encoder consists of three standard Vision Transformer layers. A patch embedding layer first partitions the input volume into non-overlapping patches with $8\times8\times8$ resolution and utilizes a convolutional layer to project them into a 768-dimensional embedding space. Each subsequent ViT layer employs a multi-head self-attention mechanism with eight attention heads to extract high-level features. For both encoder architectures, the resulting feature representations are flattened and passed through two fully connected (FC) layers to parameterize the mean ($\mu$) and variance ($\sigma$) vectors of the modality-specific latent distributions. The modality-specific features are subsequently parameterized to approximate posterior distributions and fused in the latent space using either a Product-of-Experts (PoE) or Mixture-of-Experts (MoE) strategy. After a patch embedding layer is employed to project the resulting fused CT-dose latent representation from the dimension of $512$ to $768$, it is then processed by a diffusion transformer to learn a robust multimodal representation. The DiT consists of 12 DiT blocks and models the joint structure of the fused multimodal latent distribution through the diffusion process.

During self-supervised pre-training, the encoders, latent fusion module, and DiT are jointly optimized without access to ground-truth RP labels. This enables MM-DiT to learn disease-agnostic anatomical and dosimetric representations directly from the imaging and treatment data while reducing its dependence on potentially noisy or heterogeneous toxicity labels. The self-supervised pre-training objective consists of three components. First, a diffusion denoising loss guides the DiT to learn the underlying structure of the multimodal latent space by reconstructing clean latent representations from noisy latent samples. Second, a modality reconstruction loss encourages preservation of modality-specific information and cross-modal alignment, such that information from one modality can help preserve or recover the representation of the other modality. This loss is computed as the mean squared error (MSE) between the latent features produced by the modality encoder and those reconstructed by the corresponding decoder. Each MLP-based decoder consists of two linear layers with a GELU activation between them, with an input and hidden dimension of 512. Third, a Kullback--Leibler (KL) divergence loss regularizes the fused PoE/MoE latent distribution toward a well-behaved continuous latent space, thereby facilitating stable diffusion modeling and improving the quality of the latent representation.

\textbf{Supervised Bayesian Fine-tuning.} Following self-supervised pre-training, the pretrained encoders, PoE/MoE fusion module, and DiT are retained as a multi-modal backbone and incorporated with a Bayesian classification head for downstream binary RP prediction. During supervised fine-tuning, the pre-trained backbone generates multimodal representations that are used together with binary RP labels (no RP grade $\geq$2 vs. RP grade $\geq$2) to optimize the Bayesian classifier. Monte Carlo (MC) Dropout is incorporated into the classification head to approximate Bayesian inference.

All experiments were implemented in PyTorch and conducted on an NVIDIA RTX A6000 GPU. A batch size of 2 was used for both self-supervised pre-training and supervised fine-tuning. During self-supervised pre-training, the model was trained for 500 epochs using the Adam optimizer with an initial learning rate of $1\times10^{-4}$. The learning rate was gradually reduced to $1\times10^{-5}$ during training. For supervised Bayesian fine-tuning, a two-stage optimization strategy was adopted. First, the classification framework was warm-started for 10 epochs using the Adam optimizer with an initial learning rate of $1\times10^{-4}$. To mitigate the effects of class imbalance, a weighted binary cross-entropy loss with logits was used during this warm-up stage. The framework was subsequently fine-tuned for an additional 100 epochs using a focal loss to emphasize difficult-to-classify positive RP samples. The initial learning rate was set to $1\times10^{-4}$ and decreased using a cosine annealing learning-rate scheduler. The details of the model architecture are described in the supplementary material.

\subsection{RP Risk Prediction and Model Evaluation}
During inference after supervised fine-tuning, stochastic forward passes are performed for 100 times ($T=100$) independently with dropout enabled for each input subject, producing a predictive distribution of RP risk estimates rather than a single deterministic estimate. In each forward pass, the model bypasses the label-noise transition layer and directly generates the logit corresponding to the underlying true RP outcome. Each logit is then converted to a predicted probability using a sigmoid function, yielding a set of stochastic predictions $\{p^{(t)}\}_{t=1}^{T}$. The final RP risk prediction $\hat{p}$ is defined as the predictive mean across the $T$ stochastic predictions:
\begin{align}
	\hat{p} = \frac{1}{T} \sum_{t=1}^{T} p^{(t)}.
\end{align}
To derive a discrete classification for RP grades, a decision threshold $\tau$ is applied to the mean probability $\hat{p}$, converting it into a binary label $\hat{y}$ via the indicator function:
\begin{align}
	\hat{y} = \mathds{1}[\hat{p} \geq \tau].
\end{align}
The performance of the RP risk prediction models was evaluated using four primary classification metrics: area under the receiver operating characteristic curve (AUC), accuracy, sensitivity, and specificity. Given the substantial class imbalance within the study cohorts, AUC was used as the primary measure of discrimination because it provides a threshold-independent assessment of the model's ability to distinguish patients who developed RP from those who did not. Accuracy, sensitivity, and specificity were additionally reported to provide a comprehensive assessment of classification performance at the prespecified decision threshold. For statistical comparison, paired DeLong tests were used to assess differences in AUC between MM-DiT and the other models based on predictions obtained from the same patients. When multiple pairwise comparisons were performed, the resulting $p$-values were adjusted using the false discovery rate (FDR) correction. Statistical significance was defined as an FDR-adjusted $p$-value of $p_{\mathrm{adj}} < 0.01$ or $p_{\mathrm{adj}} < 0.001$.

Model calibration was assessed using the expected calibration error (ECE), which quantifies the agreement between predicted RP probabilities and the corresponding observed outcome frequencies. Lower ECE values indicate better calibration, with an ECE of zero representing perfect agreement between predicted probabilities and observed event rates.

\subsection{Uncertainty Quantification and Decomposition}
The reliability of the RP risk predictions is assessed by quantifying the predictive uncertainty. The total predictive uncertainty $U_{\text{RP}}$ is decomposed into three components to isolate their underlying source: \textbf{aleatoric uncertainty} ($U_{\text{alea}}$), \textbf{epistemic uncertainty} ($U_{\text{epst}}$), and \textbf{label uncertainty} ($U_{\text{label}}$):
\begin{align}
	U_{\text{RP}} = U_{\text{alea}} + U_{\text{epst}} + U_{\text{label}}.
\end{align}
This decomposition provides clinically actionable insights by distinguishing between uncertainty arising from inherent data noise, model limitations, and unreliable ground-truth annotations. By quantifying these metrics separately, the framework offers a more transparent profile of prediction reliability, helping clinicians identify whether an ambiguous result stems from a lack of training data, poor image quality, or label errors.

\textbf{Aleatoric uncertainty} ($U_{alea}$) captures inherent stochasticity of the input data and its impact on model predictions. Because this uncertainty is irreducible, a high $U_{\text{alea}}$ suggests that the toxicity outcome is intrinsically ambiguous regardless of model optimization. For our binary classification task, we quantify aleatoric uncertainty by averaging the variances of the Bernoulli-distributed probabilities across $T$ stochastic MC Dropout passes \cite{nair2020exploring,herzog2020integrating}:
\begin{align}
	U_{alea}==\frac{1}{T} \sum_{t=1}^{T} p^{(t)} \big(1 - p^{(t)}\big).
\end{align}
Here, the term $p^{(t)} \big(1 - p^{(t)}\big)$ represents the closed-form variance of the Bernoulli distribution for each stochastic iteration. This metric effectively identifies cases where the underlying data lacks sufficient discriminative features to produce a certain prediction.

\textbf{Epistemic uncertainty} ($U_{epst}$), also known as model uncertainty, reflects the limitations in the model's parameters due to insufficient or non-representative training data. In clinical settings, high $U_{\text{epst}}$ typically indicates that a patient possesses characteristics such as rare anatomical features or uncommon treatment patterns that fall outside the distribution of the training cohort. For these out-of-distribution cases, the model lacks the requisite knowledge to provide a reliable prediction, suggesting they would benefit from additional clinical review or further data acquisition. We estimated epistemic uncertainty by calculating the standard deviation of the probabilistic predictions across $T$ stochastic MC Dropout passes \cite{herzog2020integrating}: 
\begin{align}
	U_{\text{epst}} = \sqrt{\frac{1}{T} \sum_{t=1}^{T} \left(p^{(t)} - \hat{p}\right)^2}.
\end{align}
Unlike aleatoric uncertainty, epistemic uncertainty is reducible; it diminishes as the model is exposed to a larger and more diverse set of training samples.

\textbf{Label uncertainty} ($U_{\text{label}}$) explicitly models inconsistencies and subjectivity in toxicity annotations. A high $U_{\text{label}}$ identifies cases where the clinical grading may be ambiguous or prone to inter-observer variability, signaling that the prediction require cautious interpretation. Label uncertainty is modeled using an additional label-noise transition layer, which maps the Bayesian classifier predictions to an estimated label-noise transition matrix and thereby accounts for uncertainty associated with potentially noisy or ambiguous RP labels. Thus, the label uncertainty is quantified as the variance induced by this transition process, capturing the expected volatility of the label given the predicted risk and estimated noise rates:
\begin{align}
	U_{\text{label}} =\hat{p} \cdot \alpha (1-\alpha) + (1 - \hat{p}) \cdot \beta (1-\beta).
\end{align}
where $\alpha$ and $\beta$ are false negative rate and false positive rate, respectively, estimated by the label-noise transition layer. This approach allows the model to remain robust even when trained on noisy clinical endpoints, effectively 'de-noising' the learning process while flagging unreliable annotations.

The decomposition of toxicity uncertainty is particularly important for toxicity prediction, where uncertainty is not solely due to model limitations but is also influenced by inherent outcome variability and subjective clinical labeling. By decomposing total predictive uncertainty into these three components, the proposed framework provides a comprehensive characterization of prediction reliability. This enables distinguishing whether uncertainty arises from model limitations, intrinsic patient variability, or annotation inconsistency. Such differentiation is particularly important in clinical practice, where decisions must account not only for predicted risk but also for the confidence in those predictions.

\section*{Acknowledgment}
This work was partially supported by the Radiation Oncology Institute (Grant \#ROI 2022-9132) and the HESI THRIVE Initiative.

\section*{Author Contributions}
Concept and design: M.C., J.Y. Data collection and curation: M.C., J.Y., T.L., J.W., J.B., S.C.T., R.V. Model development, data and statistical analysis: J.Y., M.C. Interpretation of data: M.C., J.Y., T.L., R.M.S., K.E.R. Drafting of manuscript: J.Y., T.L., M.C. Critical review and revision of the manuscript: all authors. Funding: M.C. All authors have read and approved the manuscript.

\section*{Conflict of interests}
The authors declare no conflicts of interest.

\section*{Data availability statement}
The data used in this study were obtained from two sources. The first dataset was obtained from the National Cancer Institute (NCI) National Clinical Trials Network (NCTN)/NCI Community Oncology Research Program (NCORP) Data Archive. These data were originally collected as part of the clinical trial NCT00533949, entitled “A Randomized Phase III Comparison of Standard-Dose (60 Gy) Versus High-Dose (74 Gy) Conformal Radiotherapy With Concurrent and Consolidation Carboplatin/Paclitaxel +/- Cetuximab (IND \#103444) in Patients With Stage IIIA/IIIB Non-Small Cell Lung Cancer.” Access to these data is subject to the NCTN/NCORP Data Archive's access policies and procedures. The second dataset was obtained from the Mount Sinai Health System (MSHS). These data are not publicly available due to institutional data access and privacy restrictions. Access requests must be submitted to and approved by MSHS.

\bibliography{sn-references}

\end{document}